\documentclass[11pt, a4paper]{article}
\pdfoutput=1
\usepackage{jcappub}

\usepackage{booktabs}
\usepackage{multirow}
\usepackage{amssymb}
\usepackage[export]{adjustbox}
\usepackage{floatrow}
\newfloatcommand{capbtabbox}{table}[][3.5in]
\newfloatcommand{capbfigbox}{figure}[][2.3in]
\usepackage{blindtext}
\usepackage{amsmath}
\usepackage{booktabs}
\usepackage{aas_macros}

\newcommand{\ie}{i.e.~}

\def\lsim{\mathrel{\raise.3ex\hbox{$<$\kern-.75em\lower1ex\hbox{$\sim$}}}}
\def\gsim{\mathrel{\raise.3ex\hbox{$>$\kern-.75em\lower1ex\hbox{$\sim$}}}}

\begin{document}

\hspace*{110mm}{\large \tt FERMILAB-PUB-17-568-A}\\
\hspace*{118.6mm}{\large \tt LA-UR-17-31003}
\vskip 0.2in

\title{Novel Gamma-Ray Signatures of PeV-Scale Dark Matter}


\author{Carlos Blanco,$^{a,b}$}\note{ORCID: http://orcid.org/0000-0001-8971-834X}
\emailAdd{carlosblanco2718@uchicago.edu}
\author{J. Patrick Harding$^c$}\note{ORCID: http://orcid.org/0000-0001-9844-2648}
\emailAdd{jpharding@lanl.gov}
\author{and Dan Hooper$^{b,d,e}$}\note{ORCID: http://orcid.org/0000-0001-8837-4127}
\emailAdd{dhooper@fnal.gov}

\affiliation[a]{University of Chicago, Department of Physics, Chicago, IL USA}
\affiliation[b]{University of Chicago, Kavli Institute for Cosmological Physics, Chicago, IL US}
\affiliation[c]{Physics Division, Los Alamos National Laboratory, Los Alamos, NM, USA}
\affiliation[d]{Fermi National Accelerator Laboratory, Center for Particle Astrophysics, Batavia, IL USA}
\affiliation[e]{University of Chicago, Department of Astronomy and Astrophysics, Chicago, IL USA}

\abstract{The gamma-ray annihilation and decay products of very heavy dark matter particles can undergo attenuation through pair production, leading to the development of electromagnetic cascades. This has a significant impact not only on the spectral shape of the gamma-ray signal, but also on the angular distribution of the observed photons. Such phenomena are particularly important in light of the new HAWC experiment, which provides unprecedented sensitivity to multi-TeV photons and thus to very heavy dark matter particles. In this study, we focus on dark matter in the 100 TeV-100 PeV mass range, and calculate the spectral and angular distribution of gamma-rays from dwarf galaxies and from nearby galaxy clusters in this class of models.}

\maketitle

\section{Introduction}

One of the most important outstanding questions in modern cosmology is the nature of dark matter.  Evidence of dark matter's existence comes from a wide variety of observations, including the rotational speeds of galaxies, the velocities of galaxies within clusters, gravitational lensing, the cosmic microwave background, the light element abundances, and large scale structure. But despite these many observational indications of dark matter, we remain ignorant of the particle nature of this substance. To reveal the identity of dark matter, it will be crucial to measure its non-gravitational interactions. Among the many efforts in this direction are indirect detection experiments, which aim to detect and identify the annihilation or decay products of dark matter particles.

Indirect searches for dark matter currently employ a wide range of strategies, aiming to detect photons, cosmic rays, and neutrinos from a number of different astrophysical targets. Among the targets of dark matter searches using gamma-ray telescopes are dwarf spheroidal galaxies~\citep{Fermi-LAT:2016uux,Drlica-Wagner:2015xua,Geringer-Sameth:2014qqa,Albert:2017vtb}, the Galactic Center~\citep{TheFermi-LAT:2017vmf,Hooper:2012sr,Abeysekara:2014ffg}, galaxy clusters and groups~\citep{Lisanti:2017qlb,Ackermann:2015fdi}, and the extragalactic gamma-ray background~\citep{Ackermann:2015tah,DiMauro:2015tfa}. Each of these observational approaches offers relative advantages and disadvantages.  For example, while the Galactic Center is expected to generate a very bright flux of dark matter annihilation products, observations of dwarf spheroidal galaxies suffer from far lower astrophysical backgrounds.


In this study, we focus our attention on the gamma-ray signals from very heavy dark matter particles ($m_{\chi} \sim$~0.1-100 PeV). Dark matter in this mass range has received comparitively little attention, in part due to the long appreciated point that a thermal relic cannot be heavier than $\sim$100 TeV without freezing-out of equilibirum in the early universe to yield an abundance in excess of the measured cosmological dark matter density~\citep{Griest:1989wd}. This argument, however, is predicated on two important assumptions: 1) that the universe was dominated by radiation prior to big bang nucleosynthesis, and 2) that the dark matter was initially in thermal equilibrium with the Standard Model. In scenarios in which the early universe experienced an era of matter domination~\cite{Berlin:2016vnh,Berlin:2016gtr,Fornengo:2002db,Gelmini:2006pq,Hooper:2013nia,Kane:2015jia,Patwardhan:2015kga,Hoof:2017ibo,Bramante:2017obj,Giblin:2017wlo}, or a period of late-time inflation~\cite{Davoudiasl:2015vba,Lyth:1995ka,Boeckel:2011yj,Boeckel:2009ej}, much heavier thermal relics are possible. Alternatively, very heavy dark matter can be produced with an acceptable abundance if it was not in thermal equilibrium with the Standard Model bath in the early universe~\cite{Chung:1998zb,Chung:1998rq,Allahverdi:2010xz,Giudice:1999yt,Chung:1998ua}. On more empirical grounds, very heavy dark matter particles have also received a greater degree of interest in recent years in response to IceCube's observation of a diffuse flux of high-energy astrophysical neutrinos~\cite{Chianese:2017nwe,Sahoo:2017cqg,Bhattacharya:2017jaw,Borah:2017xgm,Cohen:2016uyg,Chianese:2016kpu,Chianese:2016smc,Dev:2016qbd,Fiorentin:2016avj,Chianese:2016tmd,Chianese:2016opp,Biondi:2015yia,Boucenna:2015tra,Berezhiani:2015fba,Anchordoqui:2015lqa,Murase:2015gea,Esmaili:2014rma,Rott:2014kfa,Zavala:2014dla,Bhattacharya:2014vwa,Bai:2013nga,Esmaili:2013gha}.

The universe is largely transparent to gamma rays with energies up to $\sim$100 GeV, and photons as energetic as $\sim$100 TeV can propagate over Galactic distance scales without experiencing significant attenuation. For this reason, most indirect detection studies targeting dwarf galaxies, the Galactic Center, and even nearby galaxy clusters, safely neglect the interactions of gamma rays experienced during their propagation. For very heavy dark matter particles, however, the gamma-ray annihilation or decay products can non-negligibly scatter with the cosmic microwave and infrared backgrounds via pair production, leading to the formation of electromagnetic cascades. This phenomena not only alters the spectral shape of the gamma-ray signal, but also the angular profile of this emission.

Indirect searches for very heavy dark matter particles are currently of particular interest due to the recent release of the first results from the High Altitude Water Cherenkov (HAWC) gamma-ray observatory. Although the first constraints on annihilating dark matter published by the HAWC Collaboration~\cite{Albert:2017vtb} were only presented for masses up to 100 TeV (see also Refs.~\cite{Abeysekara:2014ffg,Harding:2015bua,Proper:2015xya,Yapici:2017bxr,Cadena:2017jmw}), this is anticipated to be extended to 1 PeV and above as analysis technique progress~\citep{Abeysekara:2014ffg}. Another recent study~\cite{Abeysekara:2017jxs} also presented constrains on dark matter annihilation and decay based on HAWC data, but focused on the relatively nearby region of the Inner Galaxy, and neglecting any possible impact from gamma-ray attenuation or cascades. In this study, we focus on the signatures of very heavy dark matter particles annihilating or decaying in more distant targets, including galaxy clusters and dwarf galaxies. As we will demonstrate, gamma-ray attenuation and the subsequent formation of electromagnetic cascades can each play an important role in the phenomenology of indirect searches for PeV-scale dark matter in such targets.


\section{The HAWC Observatory}

HAWC is a wide field-of-view observatory, sensitive to gamma rays with energies greater than 500 GeV, and located 4100 m above sea level at Sierra Negra, Mexico. Completed in 2015, HAWC consists of 300 water Cherenkov detectors (WCDs) situated over an area of 22,000 m$^2$. Each WCD is 5 m tall and 7.3 m in diameter, filled with 200 tons of purified water and instrumented with four photomultipliers~\cite{Abeysekara:2017mjj}. When a very high-energy gamma ray hits the atmosphere, it creates an electromagnetic cascade of lower-energy particles. HAWC is designed to observe the Cherenkov light from these particles as they pass through the HAWC WCDs. 

Through relative timing between the WCDs, HAWC is able to determine the direction of an original (\ie primary) gamma-ray within $1^{\circ}$ at 500 GeV, and within $0.2^{\circ}$ above tens of TeV. The measured Cherenkov light intensity in the WCDs is used to constrain the energy of the primary gamma-ray, with an uncertainty of roughly a factor of three. The WCD design enables HAWC to observe with a duty cycle greater than 90\%, regardless of time of day or weather conditions~\cite{Abeysekara:2017mjj}. 

HAWC observes approximately 2 sr of the sky at any one time. Because HAWC is located at a latitude of 19$^{\circ}$, it is primarily sensitive to gamma-ray sources with declinations between $-26^{\circ}$ and $64^{\circ}$, with diminishing sensitivity for sources beyond this range. The HAWC sensitivity to the Galactic Center ($\delta \approx -29^{\circ}$), for example, is approximately an order of magnitude weaker than that for the Crab nebula ($\delta \approx 16^{\circ}$)~\cite{Abeysekara:2017hyn}. As the Earth rotates, HAWC is able to observe more than $2/3$ of the sky each day, and continuously collects data from its entire field-of-view, unlike imaging atmospheric Cherenkov telescopes (IACTs) which must be pointed at specific targets of observation.

\begin{figure}
\centering
\includegraphics[width=0.8\textwidth]{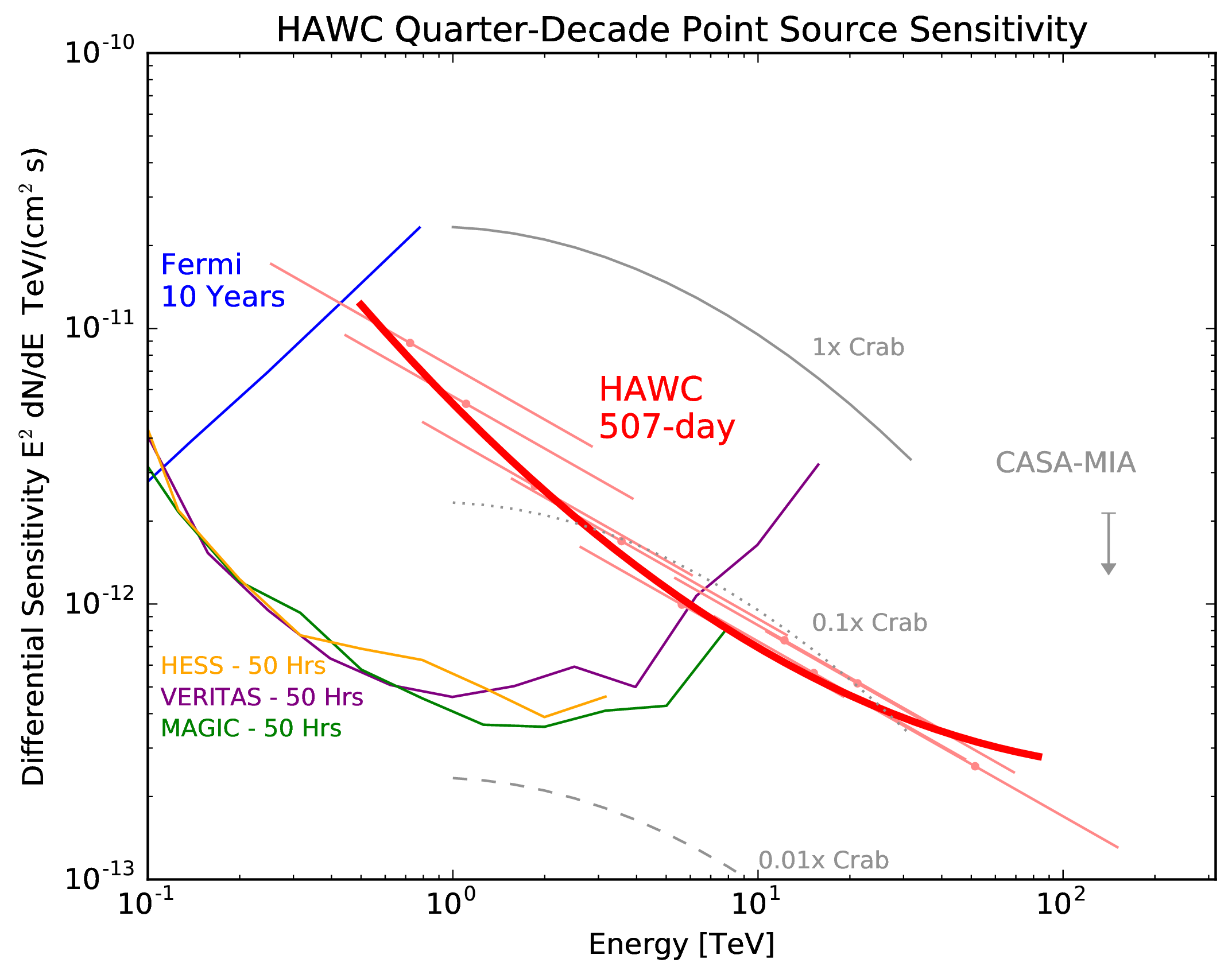}
\caption{The differential sensitivity of the HAWC observatory~\cite{Abeysekara:2017mjj} as compared to three pointed Imaging Air Cherenkov Detectors (IACTs) -- HESS~\cite{Holler:2015uca}, MAGIC~\cite{2016APh....72...76A}, and VERITAS~\cite{2015ICRC...34..771P} -- as well as the Fermi Gamma-Ray Space Telescope~\cite{Pass8}. Note that for a typical source, the mission-integrated observation time for IACTs is 10-100 hours, compared to the years of observation for either Fermi or HAWC. For more details, see Ref.~\cite{Abeysekara:2017mjj}.
\label{HAWCsens}}
\end{figure}

The primary background for HAWC consists of showers generated by hadronic cosmic rays. Although the flux of cosmic rays is much higher than that of gamma rays, HAWC is able to utilize the morphology of these events to distinguish and remove vast majority of this background, achieving an efficiency for rejection of 99.9\% at energies above 10 TeV~\cite{Abeysekara:2017mjj}. To determine the residual background flux for gamma-ray studies, the HAWC Collaboration uses a technique called ``Direct Integration'' which averages over regions of the sky within a single band of declination~\cite{2012ApJ...750...63A}. With its observed number of gamma rays and background hadrons, the HAWC sensitivity is shown in Fig.~\ref{HAWCsens}.

The sensitivity of HAWC is anticipated to greatly improve over the next year. A sparse array of smaller WCDs is currently being built around the main HAWC array~\cite{Capistran:2017fxl,Joshi:2017eou}, leading to an improvement in sensitivity of a factor of four at energies above tens of TeV. This will also have the effect of significantly reducing the energy and flux systematics for these very high-energy gamma rays. Additionally, an improved HAWC energy reconstruction effort is underway~\cite{Marinelli:2017vzu}. With the new data pass (Pass 5), which will include this improved energy reconstruction, HAWC's energy resolution is expected to improve by a factor of two.

\section{Gamma-Ray Attenuation and Electromagnetic Cascades}

The spectrum and trajectory of very high-energy gamma rays propagating through radiation and magnetic fields is continuously shaped by the production and development of electromagnetic cascades~\cite{Blanco:2017bgl,Murase:2011cy,Murase:2011yw,Murase:2012df,Murase:2012xs,Berezinsky:2016feh}. In this section, we describe these gamma-ray induced cascades and their impact on indirect searches for very heavy dark matter particles. 

\subsection{Pair Production, Inverse Compton Scattering, and the Spectrum of Electromagnetic Cascades}
\label{cascadespec}

Very high-energy gamma rays can scatter with radiation fields to produce electron-positron pairs. These charged leptons
then undergo inverse Compton scattering (ICS) with low-energy photons, generating gamma rays. Meanwhile, magnetic fields also deflect the charged leptons and induce energy losses through the production of synchrotron emission.

The optical depth for pair-production in an isotropic radiation field is given by:
\begin{eqnarray}
\tau_{\gamma \gamma}(E_{\gamma}) & = & \intop\intop\sigma_{\gamma\gamma}(E_{\gamma},\epsilon)\frac{dn(\epsilon,r)}{d\epsilon}d\epsilon dr\\
 & \approx & l\int\sigma_{\gamma\gamma}(E_{\gamma},\epsilon)\frac{dn(\epsilon)}{d\epsilon}d\epsilon,\nonumber 
\end{eqnarray}
where $E_{\gamma}$ is the energy of the incoming gamma ray, $\epsilon$ is
the energy of a target photon, $\sigma_{\gamma\gamma}$ is the total
cross section for pair-production, and $dn(\epsilon,r)/d\epsilon$ is the
differential number density of the target photons at location, $r$. If the radiation fields are assumed to be homogenous, we arrive at the second expression, where $l$ is the distance traversed. The total pair-production cross section
is approximated to within 3\% accuracy by the following expression~\cite{aharonian1983}:
\begin{eqnarray}
\sigma_{\gamma\gamma}(E_{\gamma},\epsilon) & = & \frac{3\sigma_{T}}{2s^{2}} \bigg[ \left(s-1+\frac{1}{2s}-\frac{\ln s}{2}+\ln2\right) \,  \ln\left(\sqrt{s}+\sqrt{s-1}\right) \\
 & + & \frac{\left(\ln s\right)^{2}}{8}-\frac{\left(\ln\left(\sqrt{s}+\sqrt{s-1}\right)\right)^{2}}{2}+\frac{\ln2 \, \ln s}{2}-\sqrt{s^{2}-s} \bigg],\nonumber 
\end{eqnarray}
where $s=E_{\gamma}\epsilon/m_{e}^{2}$ and $\sigma_{T}$ is the Thompson
cross section.

The differential spectrum of electrons and positrons generated in these interactions is given as follows~\cite{aharonian1983}:
\begin{equation}
\frac{dN_e}{dE_{e}}(E_e)=l\iint\frac{dN_{\gamma}}{dE_{\gamma}}\left(E_{\gamma}\right)\frac{dn}{d\epsilon}(\epsilon)\frac{d\sigma_{\gamma\gamma}}{dE_{e}}(\epsilon,E_{\gamma},E_{e}) \, d\epsilon \,dE_{\gamma}, 
\end{equation}
where $dN_{\gamma}/dE_{\gamma}$ is the spectrum of gamma rays injected from the source and $d\sigma_{\gamma\gamma}/dE_{e}$ is the differential cross section for pair production, given by:
\begin{eqnarray}
\frac{d\sigma_{\gamma\gamma}(\epsilon,E_{\gamma},E_{e})}{dE_{e}} & = & \frac{3\sigma_{T}m_{e}^{4}}{32\epsilon^{2}E_{\gamma}^{3}}\left[\frac{4E_{\gamma}^{2}}{\left(E_{\gamma}-E_{e}\right)E_{e}}\ln\left(\frac{4\epsilon E_{e}\left(E_{\gamma}-E_{e}\right)}{m_{e}^{2}E_{\gamma}}\right)-\frac{8\epsilon E_{\gamma}}{m_{e}^{2}}\right.\\
 & + & \left.\left(\frac{2E_{\gamma}^{2}\left(2\epsilon E_{\gamma}-m_{e}^{2}\right)}{\left(E_{\gamma}-E_{e}\right)E_{e}m_{e}^{2}}\right)-\left(1-\frac{m_{e}^{2}}{\epsilon E_{\gamma}}\right)\frac{E_{\gamma}^{4}}{\left(E_{\gamma}-E_{e}\right)^{2}E_{e}^{2}}\right].\nonumber 
\end{eqnarray}

\begin{figure}
\includegraphics[scale=0.45]{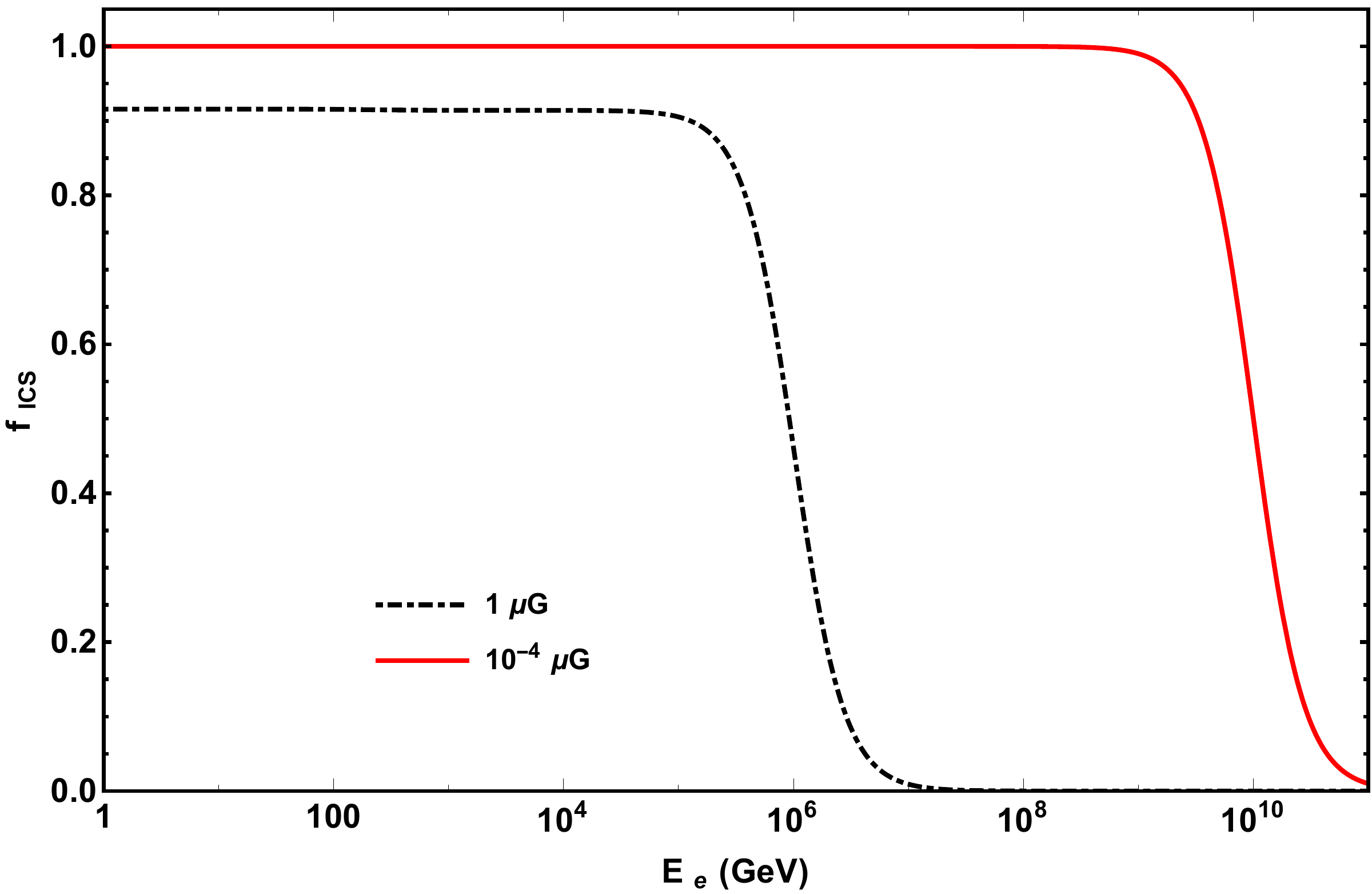} 
\caption{The fraction of electron energy losses that go into inverse Compton scattering (as opposed to synchrotron emission) for two choices of the magnetic field strength (1\,$\mu$G and $10^{-4}\,\mu$G) assuming that ICS losses are dominated by interactions with the CMB and the cosmic infrared background~\cite{Dominguez:2010bv}. At the highest energies, inverse Compton scattering is Klein-Nishina suppressed, enabling synchrotron to potentially dominate.}
\label{fics}
\end{figure}

These charged leptons will then go on to generate gamma rays through ICS, while at the same
time experiencing synchrotron cooling. Over the course of losing an amount of energy, $\Delta E_e$, an electron or positron of energy, $E_{e}$, will produce the following gamma-ray spectrum:
\begin{eqnarray}
\frac{dN_{\gamma}}{dE_{\gamma}}\left(E_{\gamma},E_{e}\right)_{\Delta E_{e}} & = & A\left(E_{e},\Delta E_{e}\right)f_{ICS}\left(E_{e}\right)l_{e}\int\frac{dn}{d\epsilon}(\epsilon)\frac{d\sigma_{ICS}}{dE_{\gamma}}(\epsilon,E_{\gamma},E_{e})d\epsilon,
\end{eqnarray}
where $A$ is a normalization factor that is set by $\Delta E_{e}=\int dE_{\gamma}E_{\gamma}dN_{\gamma}/dE_{\gamma}$, and $f_{ICS}$ is the fraction of the energy losses which go into ICS (the remaining fraction, $1-f_{ICS}$, goes into synchrotron emission). In Fig.~\ref{fics} we plot this quantity for two choices of the magnetic field strength (1\,$\mu$G and $10^{-4}\,\mu$G) assuming that ICS losses are dominated by interactions with the CMB and the cosmic infrared background~\cite{Dominguez:2010bv}. Note that inverse Compton scattering is Klein-Nishina suppressed at the highest energies, enabling synchrotron to dominate.

The differential cross section for ICS is given by \citep{aharonian1981}: 
\begin{eqnarray}
\frac{d\sigma_{ICS}}{dE_{\gamma}}(\epsilon,E_{\gamma},E_{e}) & = & \frac{3\sigma_{T}m_{e}^{2}}{4\epsilon E_{e}^{2}}\left[1+\left(\frac{z^{2}}{2\left(1-z\right)}\right)+\left(\frac{z}{\beta\left(1-z\right)}\right)-\left(\frac{2z^{2}}{\beta^{2}\left(1-z\right)}\right)\right.\\
 & - & \left(\frac{z^{3}}{2\beta\left(1-z\right)^{2}}\right)-\left(\frac{2z}{\beta\left(1-z\right)}\right)\ln\left(\frac{\beta\left(1-z\right)}{z}\right) \bigg], \nonumber 
\end{eqnarray}
where $\beta\equiv4\epsilon E_{e}/m_{e}^{2}$ and $z\equiv E_{\gamma}/E_{e}$.
The total spectrum, $dN_{\gamma}/dE_{\gamma}$, of photons in a cascade created by an electron of energy, $E_{e}$, is then calculated by taking a sum over the spectra generated as the charged lepton loses its energy
through successive scatterings.

\subsection{The Angular Distribution of Gamma-Rays in Electromagnetic Cascades}
\label{angle}

In traveling a distance, $l$, through a uniform magnetic field, an electron is deflected by the following angle~\cite{Neronov:2009gh,Tashiro:2013bxa,chen2015search}:
\begin{equation}
\sin \Theta \simeq l/r_{l}\sim0.3\;\left(\frac{B}{10^{-11}\,\rm{G}}\right)\left(\frac{100\;\mathrm{TeV}}{E_{e}}\right)^{2},
\end{equation}
where $r_L$ is the Larmor radius of the particle. If $\Theta \gsim 1$, the direction of the ICS photon
emission is essentially isotropic with respect to the original gamma ray. 

The spatial distribution of pair production events is described as follows:
\begin{eqnarray}
\frac{dP}{dV}(E_{\gamma}) &=& -\frac{d (e^{-r/\lambda(E_{\gamma})})}{dr} \\
&=& \frac{1}{4\pi r^2 \lambda(E_{\gamma})} e^{-r/\lambda(E_{\gamma})}, \nonumber
\end{eqnarray}
where $\lambda(E_{\gamma}) \equiv \int d\epsilon \, n(\epsilon) \, \sigma_{\gamma \gamma}(E_{\gamma},\epsilon)$ is the mean free path and $r$ is the distance from the source. Once an electron-positron pair is produced, these particles rapidly lose their energy via ICS and synchrotron processes, during which they traverse a negligible distance. 

The gamma-ray flux (per area, time, and solid angle) observed from an angle, $\theta$, away from the source is given by the following:
\begin{eqnarray}
F_{\gamma}(\theta)_{\rm obs} = F_{\gamma, {\rm pair}} \times \bigg[\frac{1}{4\pi \lambda(E_{\gamma})}\int_{los} \frac{\exp(-\sqrt{R^2+l^2-2Rl\cos \theta}/\lambda(E_{\gamma}))}{R^2+l^2-2Rl\cos \theta} dl \bigg],
\end{eqnarray}
where $F_{\gamma,{\rm pair}}$ is the gamma-ray spectrum per pair production event originating from a photon of energy, $E_{\gamma}$, $R$ is the distance to the source, and the integral is carried out over the line-of-sight.

A fraction of the gamma-rays produced in these electromagnetic cascades will go on to undergo further pair production events. To account for this, the above expression is implemented iteratively, until $\lambda$ becomes much larger than the other distance scales in the problem.

\section{Gamma-Rays From Dwarf Galaxies}

Due to their low astrophysical backgrounds and relative proximity, dwarf spheroidal galaxies represent a very attractive target for indirect dark matter searches, including those carried out by Fermi~\cite{Fermi-LAT:2016uux,Drlica-Wagner:2015xua,Geringer-Sameth:2014qqa}, HESS~\cite{Abramowski:2014tra,Abramowski:2010aa}, VERITAS~\cite{Archambault:2017wyh} and MAGIC~\cite{Ahnen:2016qkx,Aleksic:2011jx}. The HAWC Collaboration has recently presented the results of their search for dark matter annihilation or decay products from 15 of the Milky Way's dwarf galaxies, yielding the strongest current constraint on the annihilation cross section of dark matter particles heavier than approximately $\sim$10 TeV~\cite{Albert:2017vtb}. These systems are located at distances between $\sim$23 kpc (Segue I) and $\sim$250 kpc (Leo I) from the Solar System. For sources in this range of distances, gamma-rays with energies above $\sim$100 TeV can undergo pair production after scattering with the cosmic microwave background (CMB), attenuating the spectrum and leading to the formation of gamma-ray halos around each dwarf, several degrees in extent. 

Here, we focus on the example of the dwarf galaxy Draco, which is located $80\pm 7$ kpc from the Solar System and is located within HAWC's field-of-view ($\delta \approx 58^{\circ}$). The total flux of gamma-rays from this source (prior to any attenuation or electromagnetic cascades) is given by the following:
\begin{eqnarray}
\label{Jfactor}
\frac{d\Phi_{\gamma}}{dE_{\gamma}} \left( E_{\gamma} \right) & = & \frac{\left\langle \sigma v\right\rangle _{\rm ann}}{8\pi m_{\chi}^{2}}\frac{dN_{\gamma}}{dE_{\gamma}}\left(E_{\gamma} \right) \int_{\Delta \Omega}  d\Omega  \intop_{los}dl \, \rho^{2},\\
 & = & \frac{\left\langle \sigma v\right\rangle_{\rm ann}}{8\pi m_{\chi}^{2}}\frac{dN_{\gamma}}{dE_{\gamma}}\left(E_{\gamma}\right)\mathcal{J}_{tot}, \nonumber 
\end{eqnarray}
where $\langle \sigma v \rangle_{\rm ann}$ is the annihilation cross section, $m_{\chi}$ is the mass of the dark matter particle, and $dN_{\gamma}/dE_{\gamma}$ is the spectrum of gamma-rays produced per annihilation, which we calculate using PYTHIA~\cite{pythia}. The integrals are performed over the solid angle observed and over the line-of-sight (los). The dark matter density profile for a given dwarf galaxy is constrained by the stellar kinematics of the system. For the case of Draco, the authors of Ref.~\cite{Martinez:2013els} determine a value of $\log_{10} (\mathcal{J}_{\rm tot}/{\rm GeV}^{2} \, {\rm cm}^{-5})=18.8 \pm0.16$ (see also Refs.~\cite{Geringer-Sameth:2014yza,Bonnivard:2015xpq,Chiappo:2016xfs}), although this may underestimate the impact of some systematic uncertainties~\cite{Ichikawa:2016nbi,Hayashi:2016kcy,Ullio:2016kvy}.

The flux of gamma rays from decaying dark matter in a dwarf galaxy such as Draco can be similarly calculated as follows:
\begin{eqnarray}
\frac{d\Phi_{\gamma}}{dE_{\gamma}} \left( E_{\gamma} \right) & = & \frac{1}{4\pi m_{\chi} \tau_{\chi}}\frac{dN_{\gamma}}{dE_{\gamma}}\left(E_{\gamma} \right) \int_{\Delta \Omega}  d\Omega  \intop_{los}dl \, \rho,
\end{eqnarray}
where $\tau_{\chi}$ is the lifetime of the dark matter particle. Although we will present our results in terms of dark matter annihilation, they can easily be translated into the case of dark matter decay.




\begin{figure}
\includegraphics[scale=0.48]{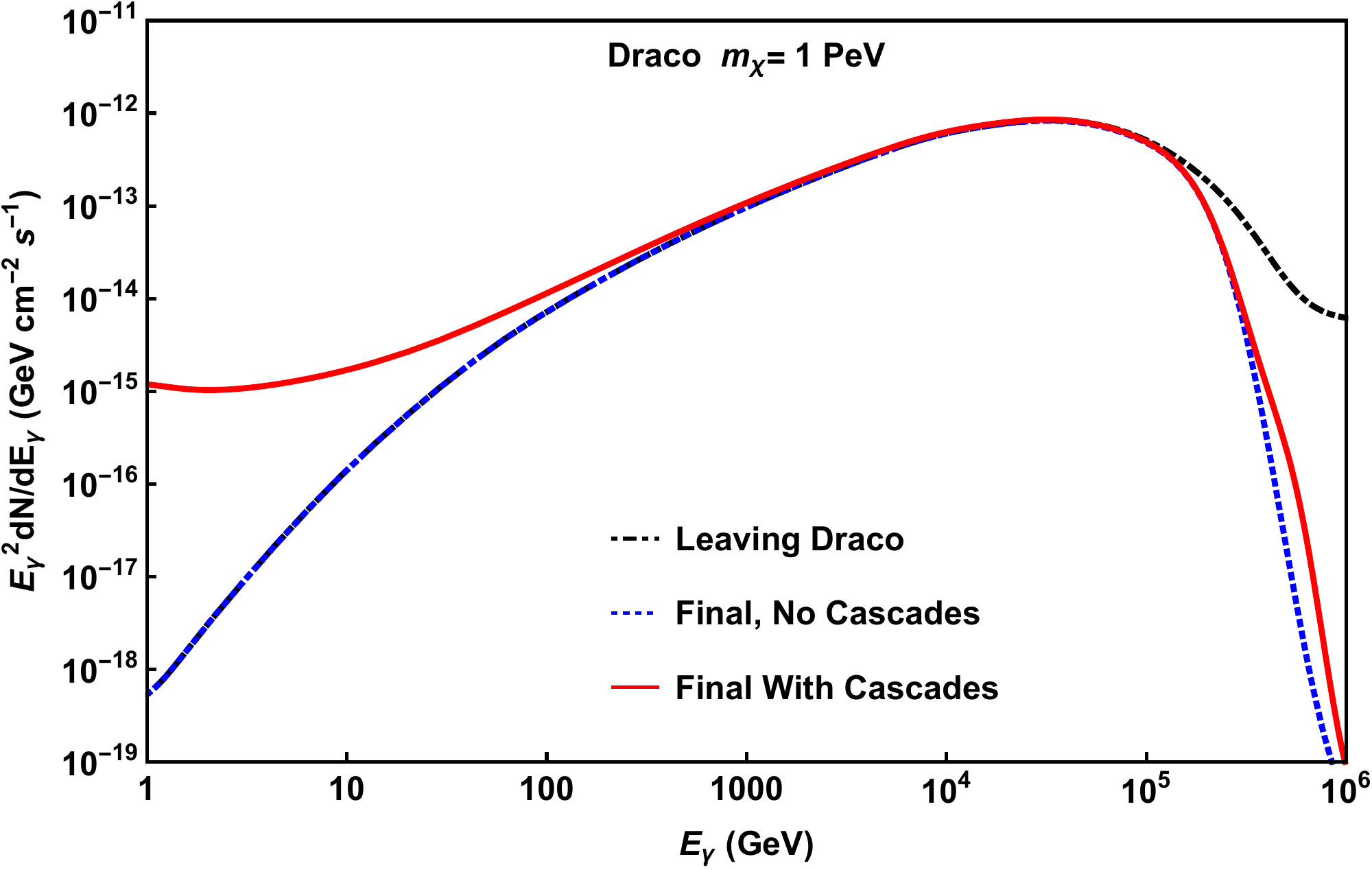} \\
\vspace{0.3cm}
\includegraphics[scale=0.48]{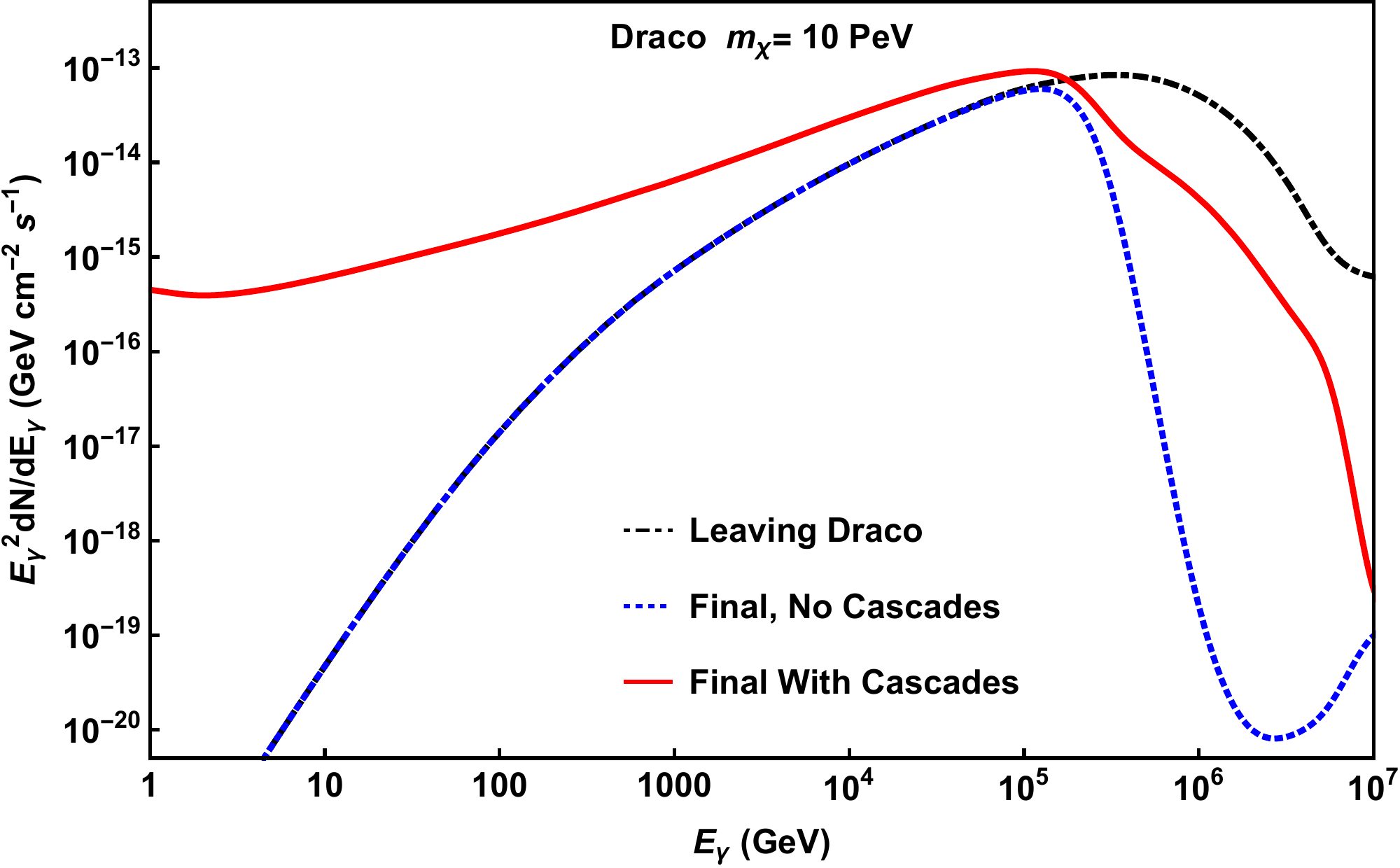} \\
\vspace{0.3cm}
\includegraphics[scale=0.48]{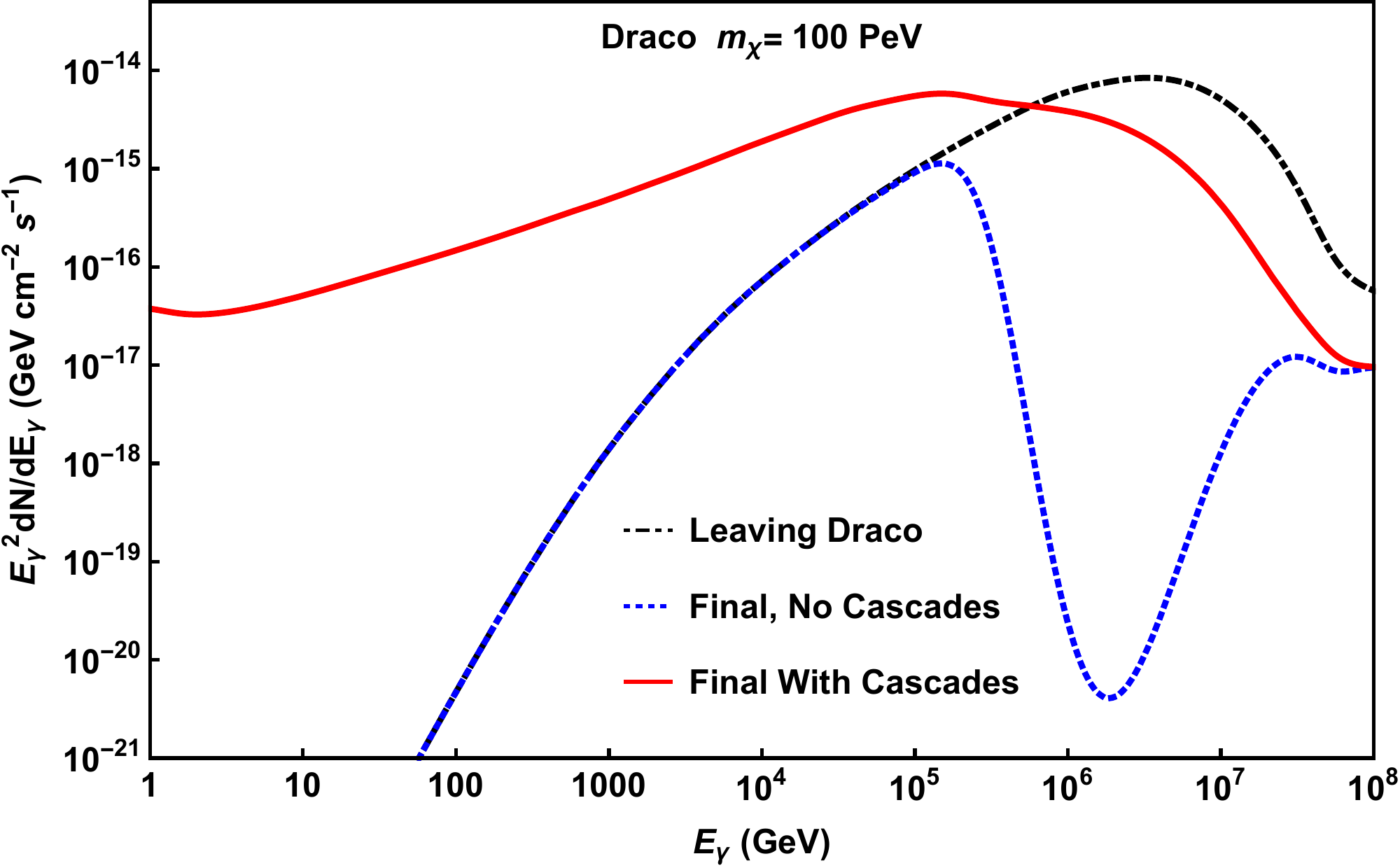}
\caption{The gamma-ray spectrum from the Draco dwarf galaxy for the case of a dark matter particle with a mass of 1, 10 or 100 PeV, annihilating to $b\bar{b}$ with a cross section of $\left\langle \sigma v\right\rangle =10^{-23} \, \rm{cm}^{3}/\rm{s}$. In each frame, the black dot-dashed curve denotes the injected spectrum, without taking into account any attenuation or cascades, while the blue dotted line represents the spectrum after attenuation. The red curve is the total spectrum, including those generated in the cascades.}
\label{dwarfspec}
\end{figure}

\begin{figure}
\includegraphics[scale=0.48]{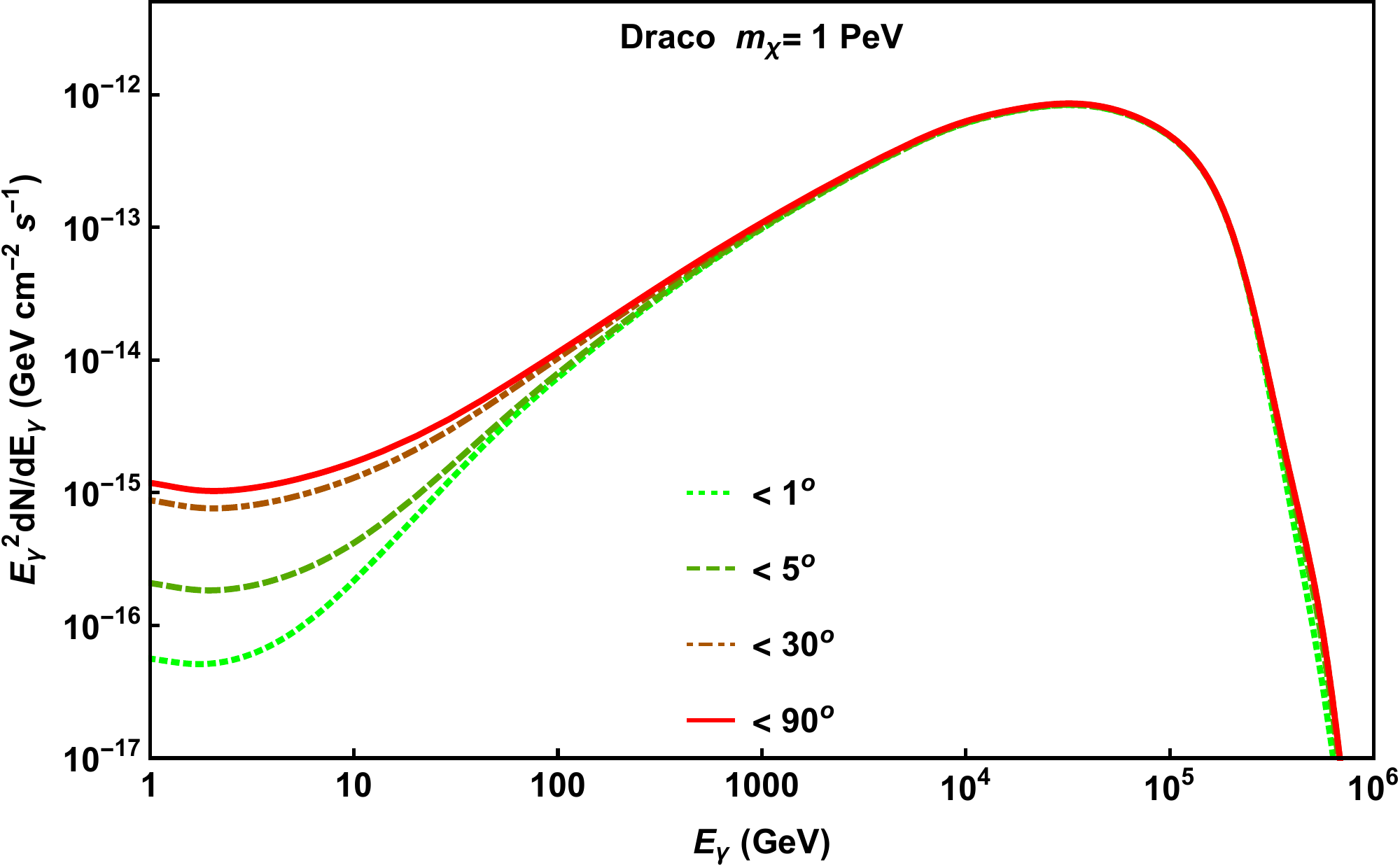}\\
\vspace{0.3cm}
\includegraphics[scale=0.48]{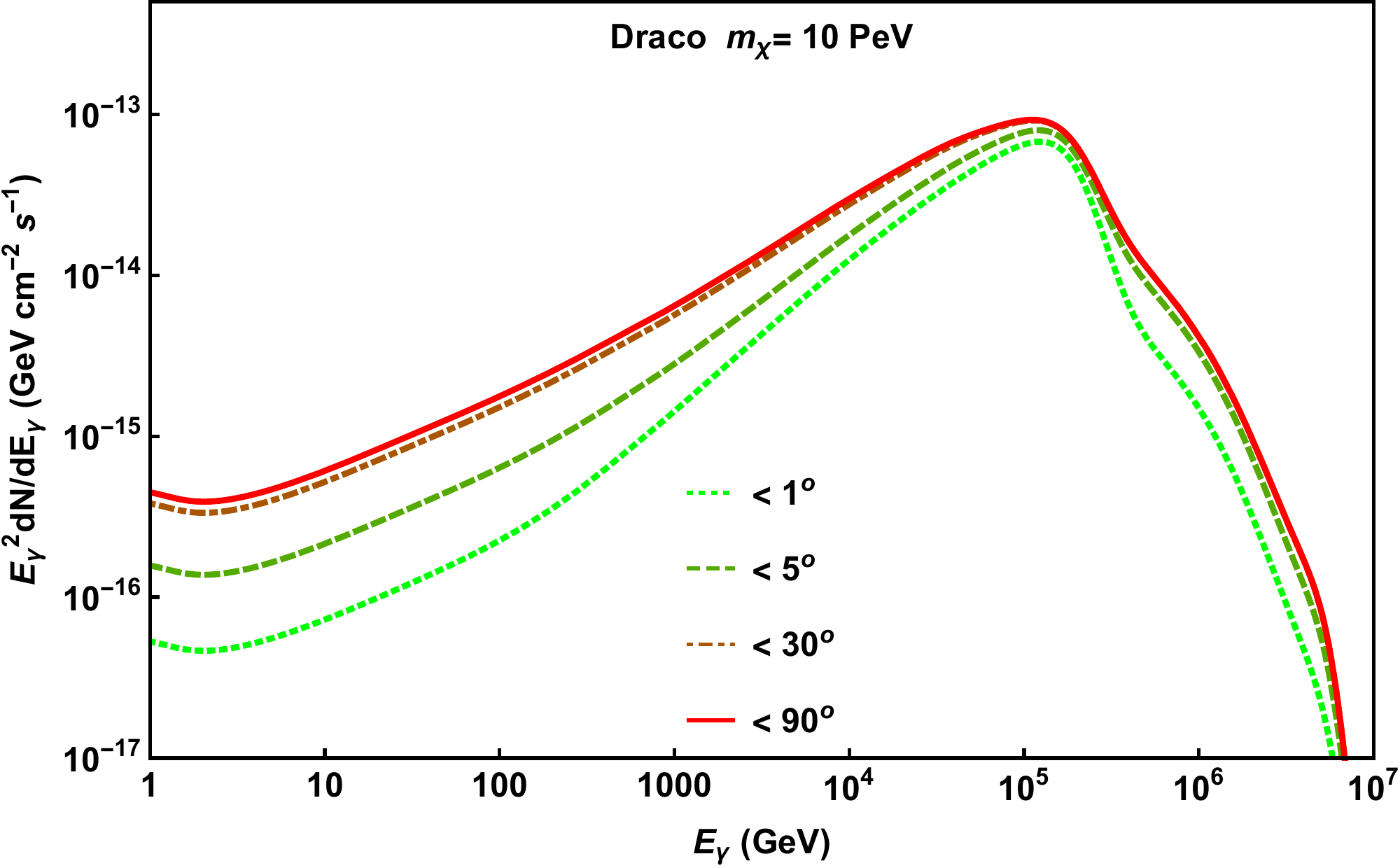} \\
\vspace{0.3cm}
\includegraphics[scale=0.48]{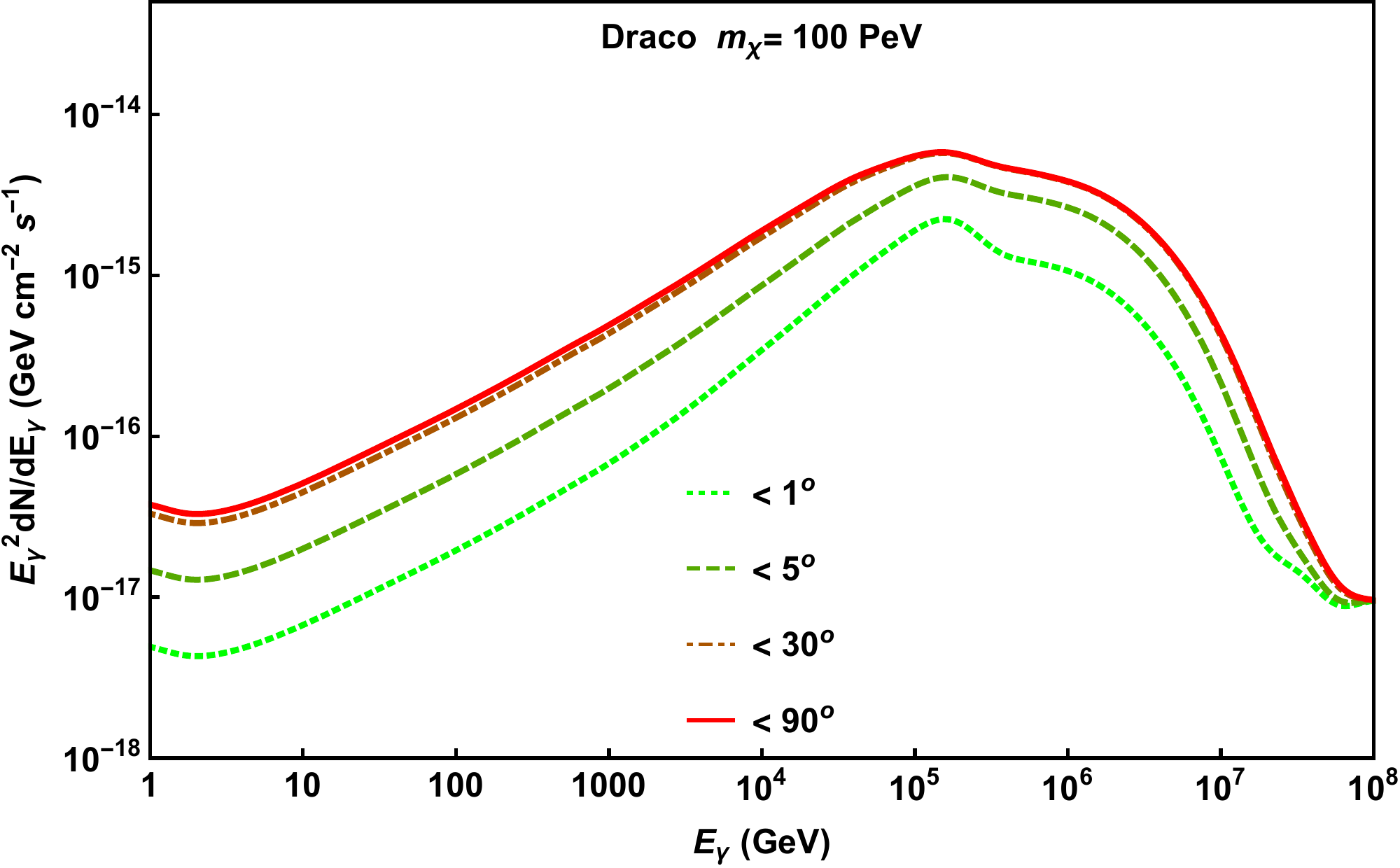}
\caption{The gamma-ray spectrum from the Draco dwarf galaxy for the case of a dark matter particle with a mass of 1, 10 or 100 PeV, annihilating to $b\bar{b}$ with a cross section of $\left\langle \sigma v\right\rangle =10^{-23} \, \rm{cm}^{3}/\rm{s}$. In each frame, we plot the spectrum integrated within a cone of radius $1^{\circ}$, $5^{\circ}$, $30^{\circ}$, or $90^{\circ}$ around Draco. Comparing this to the results shown in Fig.~\ref{dwarfspec}, we see that much of the cascade emission from this source is distributed in a region of the sky several degrees in radius.}
\label{dwarfangle}
\end{figure}

In Fig.~\ref{dwarfspec}, we plot the gamma-ray spectrum from Draco for the case of dark matter with a mass of 1, 10 or 100 PeV, annihilating to $b\bar{b}$ with a cross section of $\left\langle \sigma v\right\rangle =10^{-23} \, \rm{cm}^{3}/\rm{s}$. In each frame, the black dot-dashed curve denotes the injected spectrum, without taking into account any attenuation or cascades, while the blue dotted line represents the spectrum after attenuation. The red curve is the total spectrum, including both the unattenuated gamma-rays and those generated in the subsequent cascade. 

In calculating the formation and development of these cascades, we adopt a radiation field consisting only of the CMB and neglect energy losses from synchrotron. At energies above a couple hundred TeV, pair production with the CMB strongly attenuates the spectrum, transferring this portion of the injected energy into much lower energy cascade photons. For a 1 PeV dark matter particle, the cascade emission dominates the total spectrum at energies below several tens of GeV, whereas the cascade dominate at all energies for $m_{\chi} \gsim 10$ PeV. 

The angular distribution of the unattenuated gamma rays is determined by the square of the dark matter density profile (see Eqn.~\ref{Jfactor}), corresponding to an angular extent in the case of Draco that is comparable to the angular resolution of HAWC. In contrast, the cascade emission is further broadened by the deflection of cascade electrons in the Galactic magnetic field. Recall from Sec.~\ref{angle} that for $B \gsim 3 \times 10^{-3} \mu {\rm G}\, (E_e/{\rm PeV})^2$, the ICS emission is emitted approximately isotropically; a condition that is easily satisfied in the case of Draco. 

In Fig.~\ref{dwarfangle}, we plot the gamma-ray spectrum from Draco, as found within a cone of radius $1^{\circ}$, $5^{\circ}$, $30^{\circ}$, or $90^{\circ}$. Comparing this to the results shown in Fig.~\ref{dwarfspec}, we see that much of the cascade emission from this source is distributed in a region of the sky several degrees in radius.

\section{Gamma-Rays From Nearby Galaxy Clusters}

\begin{figure}
\includegraphics[scale=0.45]{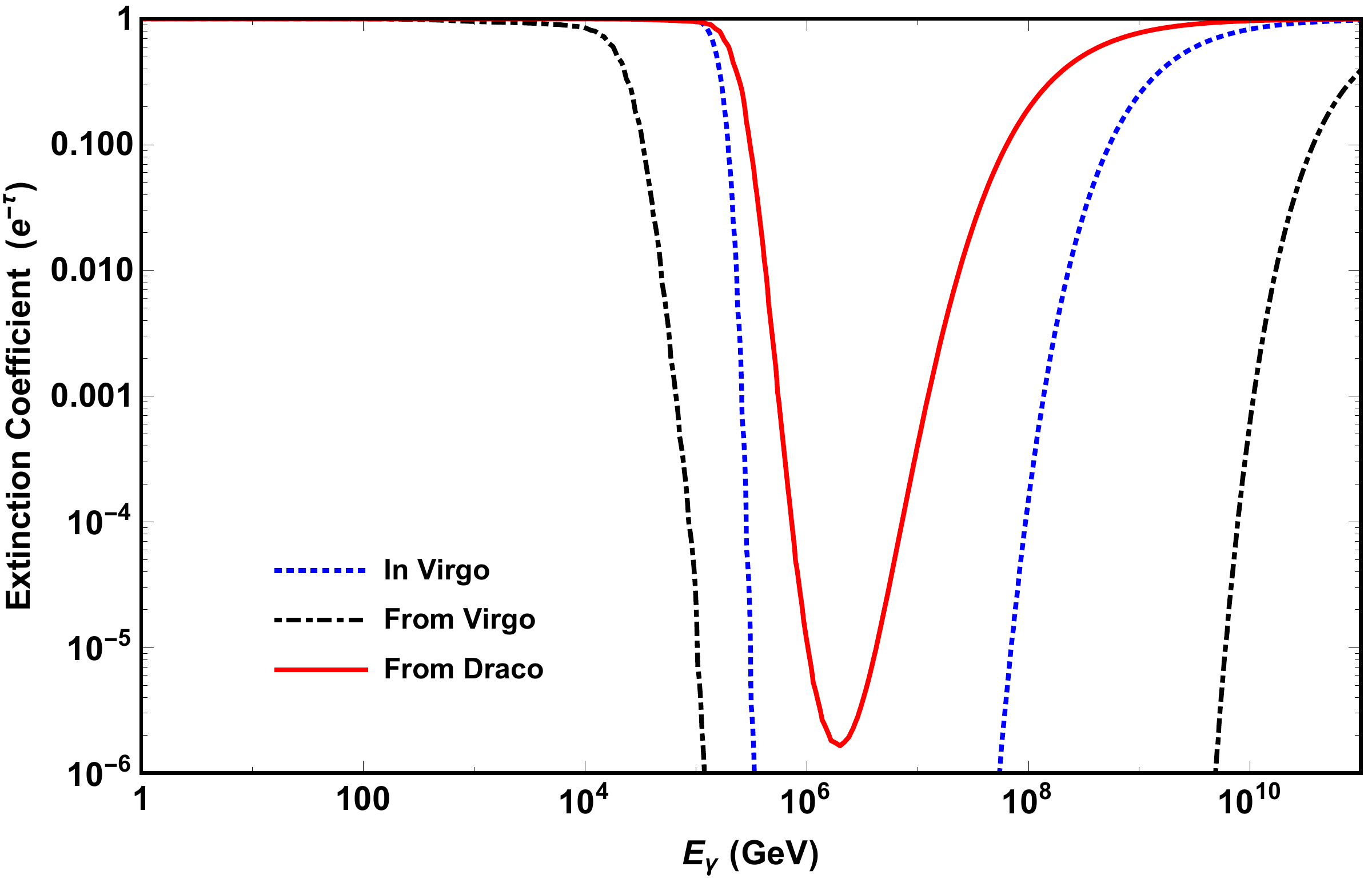} 
\caption{A comparison of the degree of attenuation predicted for a gamma ray traveling from the Draco dwarf galaxy (red solid) to that predicted to take place within the innermost 500 kpc of the Virgo Cluster (blue dotted) and along the trajectory from this source (black dot-dashed).}
\label{extinction}
\end{figure}

Galaxy clusters are also attractive targets for indirect dark matter searches, and searches for gamma-rays from such sources have been carried out by Fermi~\cite{Lisanti:2017qlb,Ackermann:2015fdi}, HESS~\cite{Abramowski:2012au}, VERITAS~\cite{Pfrommer:2012mm}, MAGIC~\cite{Aleksic:2009ir}, and HAWC~\cite{Cadena:2017jmw,Yapici:2017bxr}. In this section, we turn our attention to the effects of gamma-ray attenuation and the evolution of electromagnetic cascades on the annihilation or decay products of very heavy dark matter particles, focusing on the case of the Virgo Cluster. 

The calculation of the gamma-ray signal from a galaxy cluster differs from that of dwarf galaxies in three important ways. First, nearby galaxy clusters are themselves extended by several degrees, and cannot be treated as approximately point-like. Second, a non-negligible degree of gamma-ray attenuation is expected to take place within the cluster itself. And third, the much greater distance to the Virgo Cluster ($d\approx 16.8$ Mpc) leads to much greater attenuation, especially at energies in the range of $E_{\gamma}\sim 20-200$ TeV. In Fig.~\ref{extinction}, we compare the degree of attenuation predicted along the trajectory from the Draco dwarf galaxy to that predicted within the innermost 500 kpc of the Virgo Cluster and along the trajectory from this source.

For Virgo's dark matter density profile, we adopt a standard NFW distribution~\cite{Navarro:1995iw}:
\begin{equation}
\rho(r)=\frac{\rho_{0}}{\left(\frac{r}{r_{s}}\right)\left(1+\frac{r}{r_{s}}\right)^{2}},
\end{equation}
where $r$ is the distance to the halo's center, and we adopt $\rho_{0}=7.4\times10^{5}\,M_{\odot} \, {\rm kpc}^{-3}$ and $r_{s}=465\,{\rm kpc}$. For these parameters, the total mass of Virgo within its virial radius ($r_{v}=1825$ kpc) is $7.5\times10^{14}M_{\odot}$~\cite{han2012constraining}.


To account for the non-negligible extent of the Virgo Cluster, we follow the same approach as described in the previous section, but further convolve the angular distribution of the gamma rays by the direction-dependent $J$-factor, $\mathcal{J}\left(\Omega \right) \equiv \intop_{los}dl \, \rho^{2}(r)$.





\begin{figure}
\includegraphics[scale=0.35]{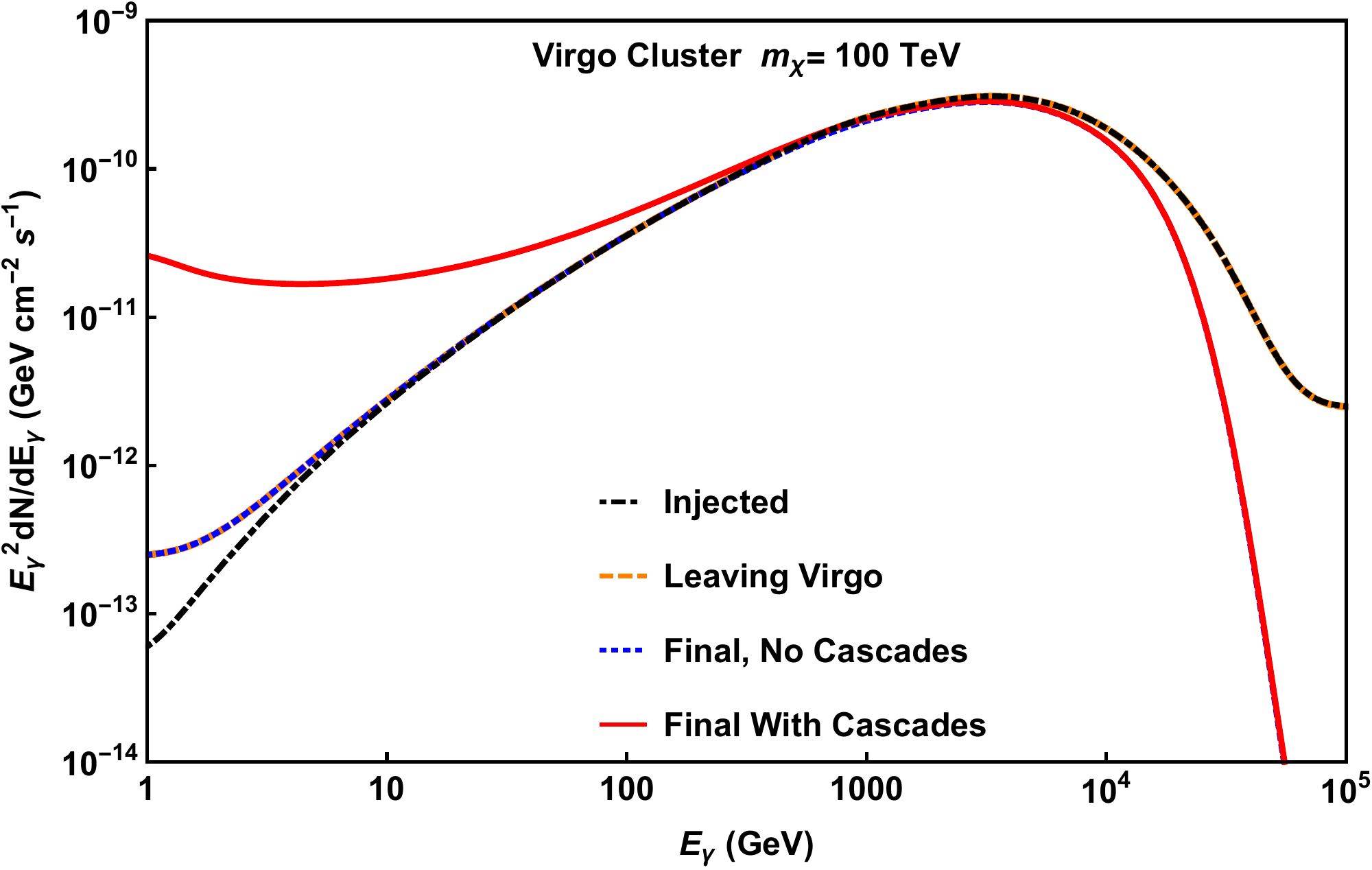}
\hspace{0.3cm}
\includegraphics[scale=0.35]{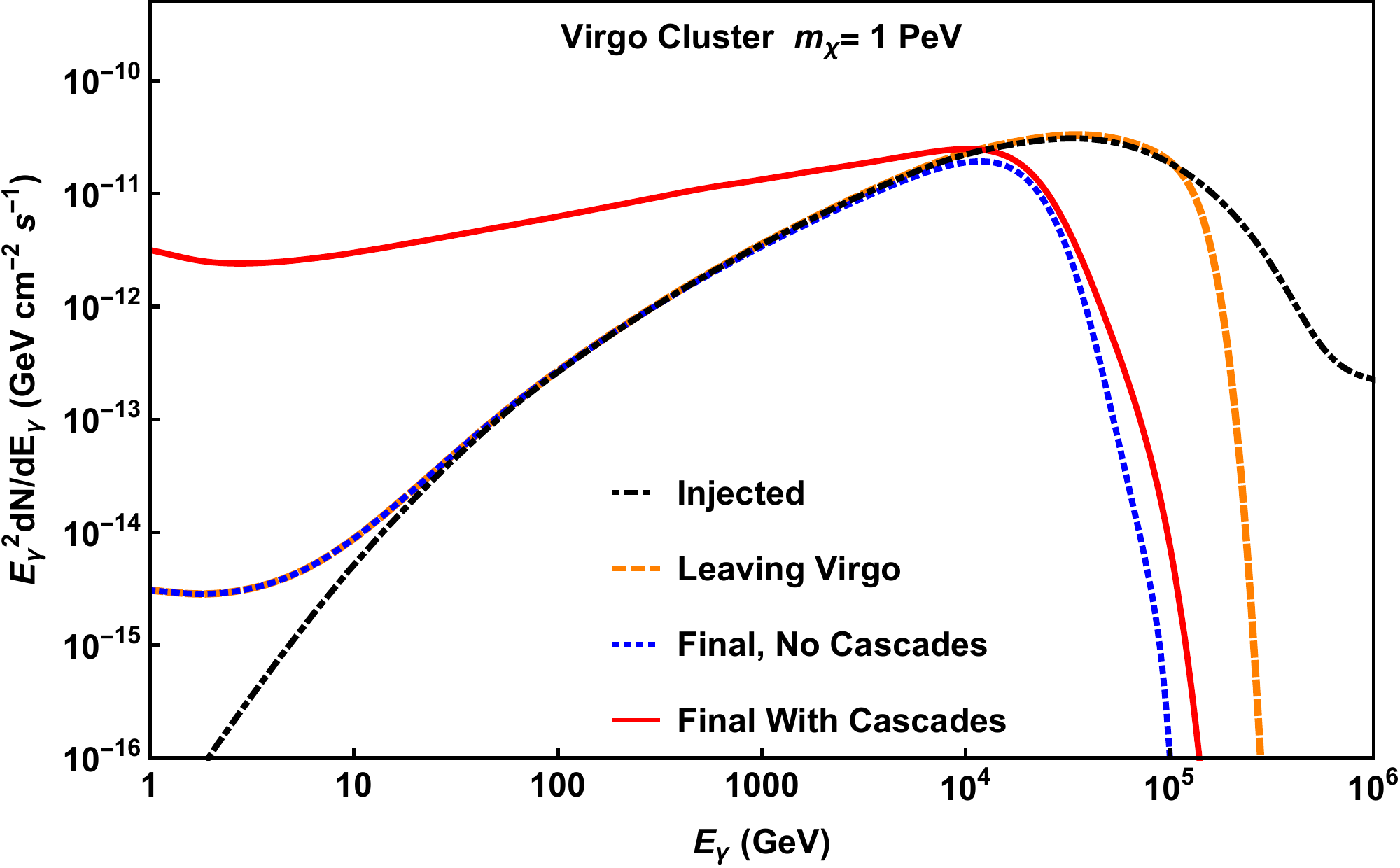}\\
\vspace{0.3cm}
\includegraphics[scale=0.35]{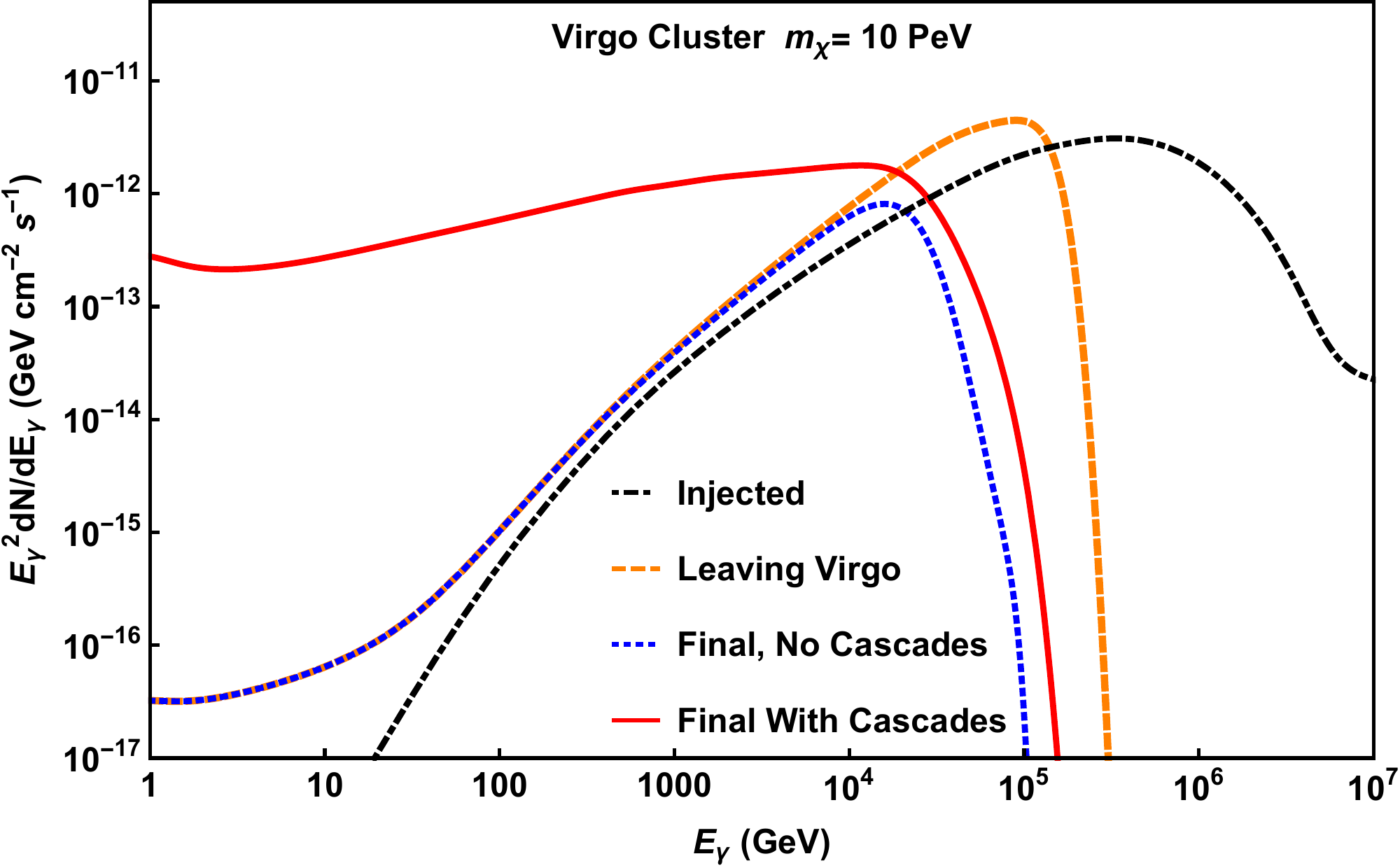}
\hspace{0.3cm}
\includegraphics[scale=0.35]{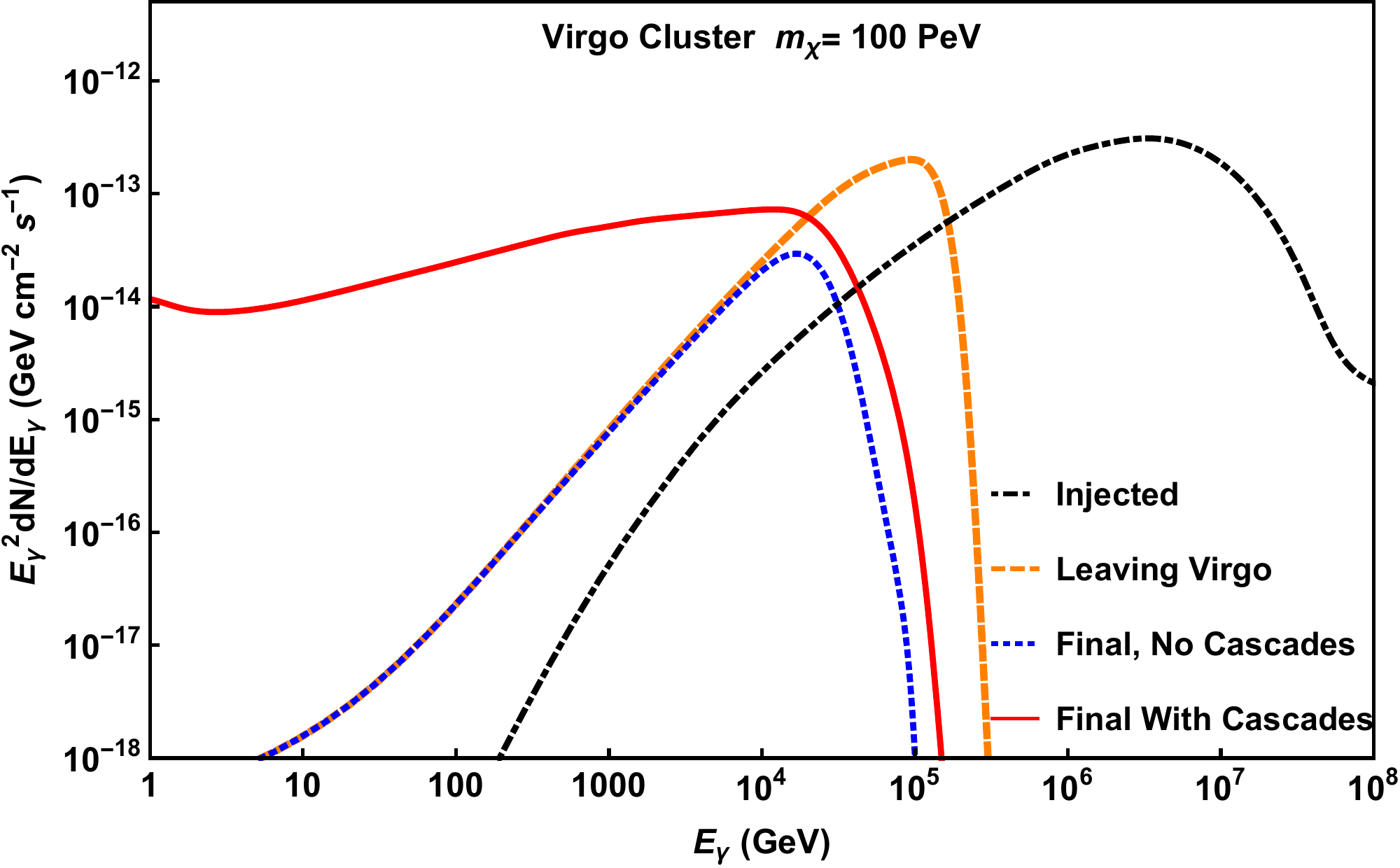}
\caption{The gamma-ray spectrum from the Virgo Cluster for a dark matter particle with mass of 0.1, 1, 10 or 100 PeV, annihilating to $b\bar{b}$ with a cross section of $\left\langle \sigma v\right\rangle =10^{-23} \, \rm{cm}^{3}/\rm{s}$. In each frame, the black dot-dashed curve denotes the injected spectrum, without taking into account any attenuation or cascades, while the dashed orange line represents the spectrum which escapes from the Virgo Cluster itself, including those generated in the cascades. The blue dotted (red solid) line is the spectrum that reaches the Solar System without (with) the contribution from cascades.}
\label{virgospec}
\end{figure}

\begin{figure}
\includegraphics[scale=0.35]{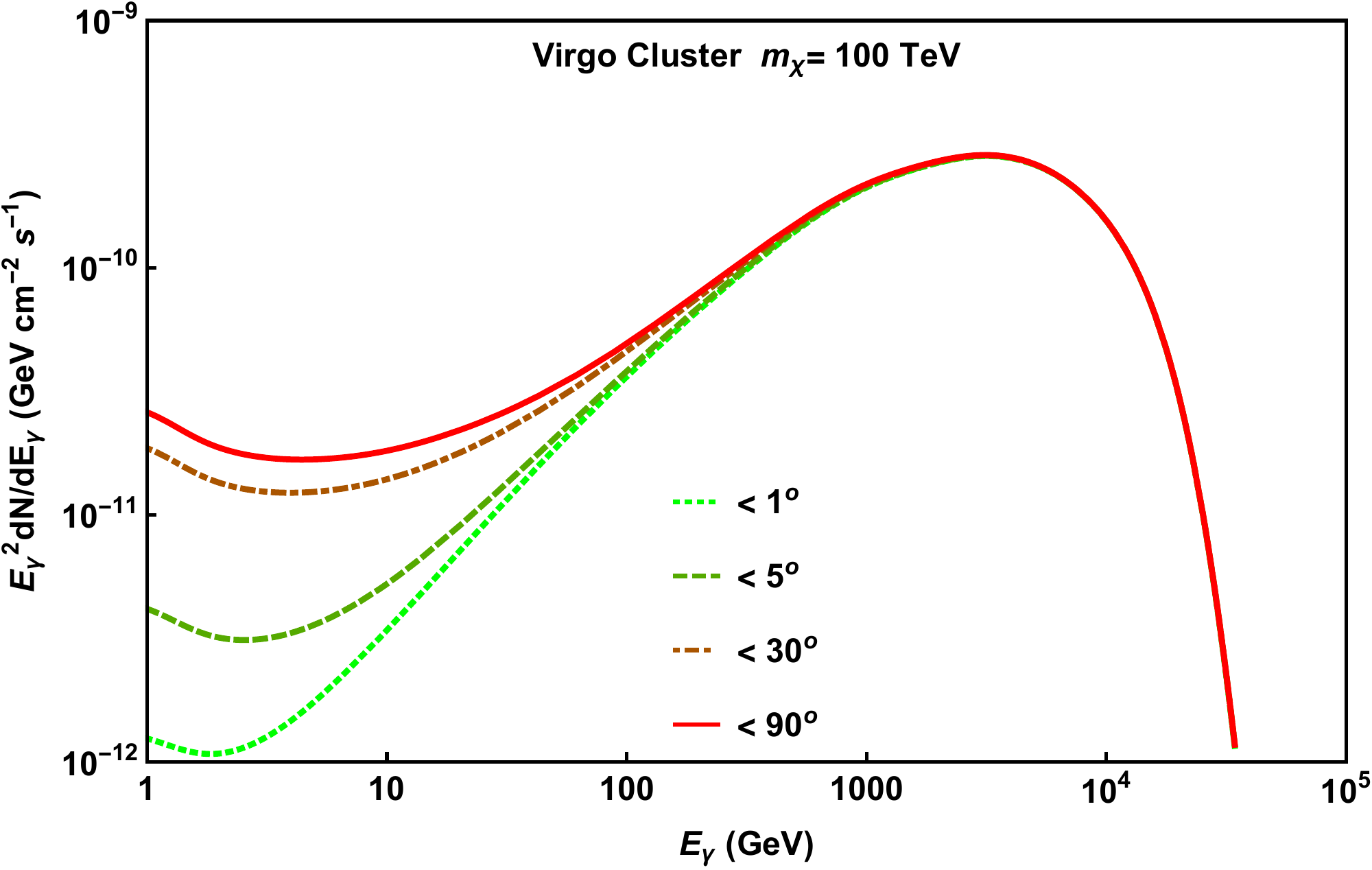}
\hspace{0.3cm}
\includegraphics[scale=0.35]{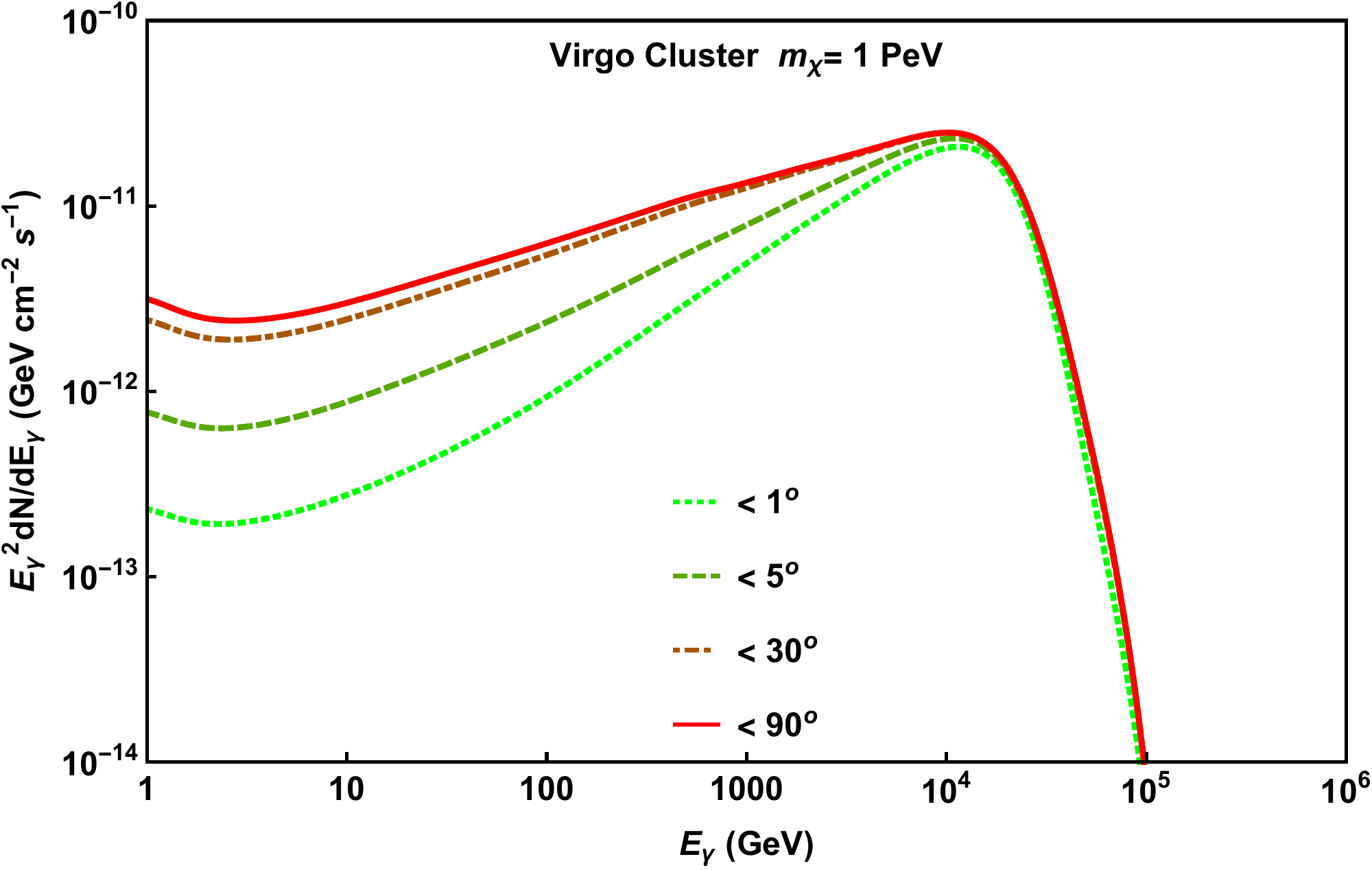}\\
\vspace{0.3cm}
\includegraphics[scale=0.35]{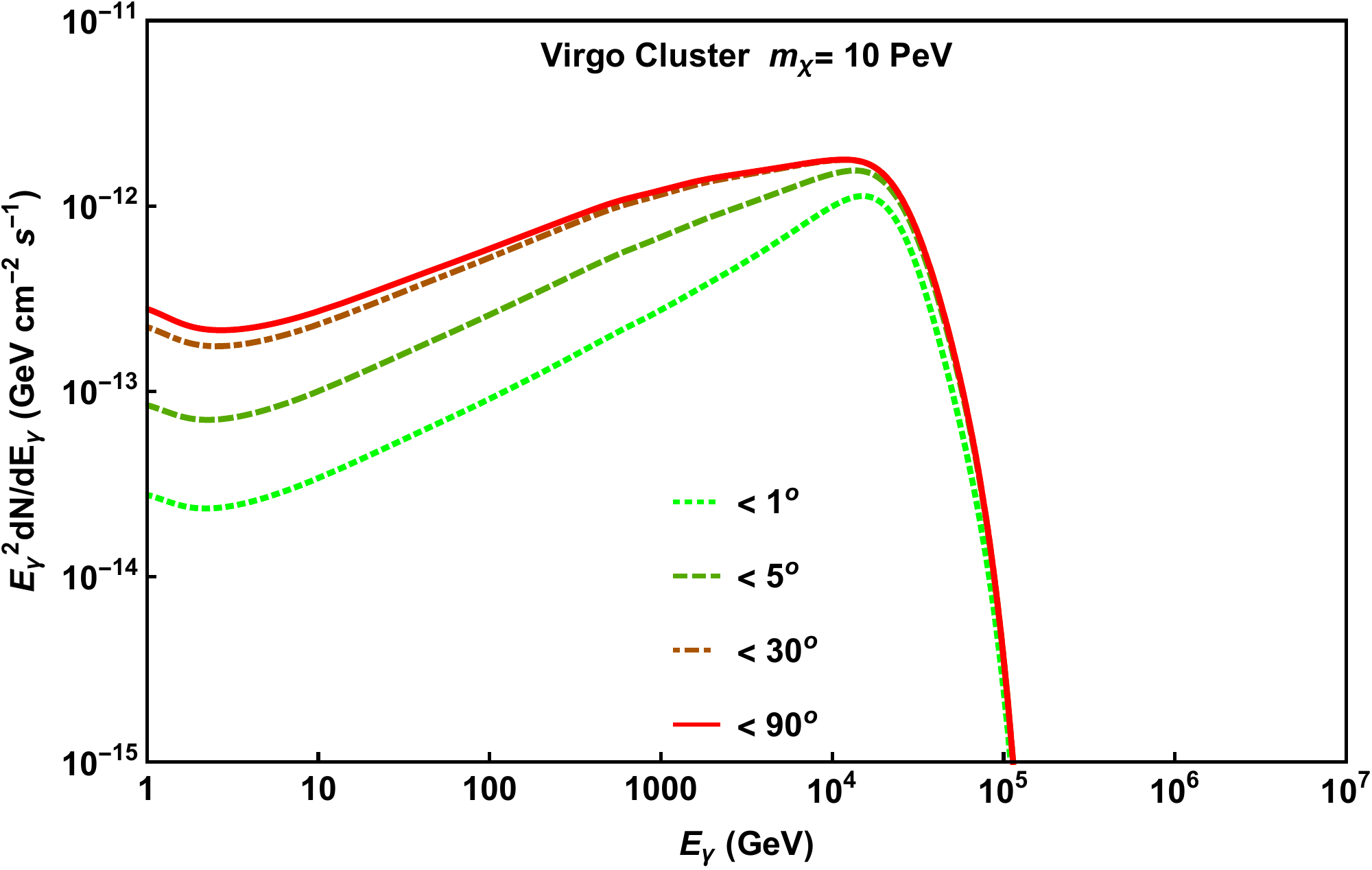}
\hspace{0.3cm}
\includegraphics[scale=0.35]{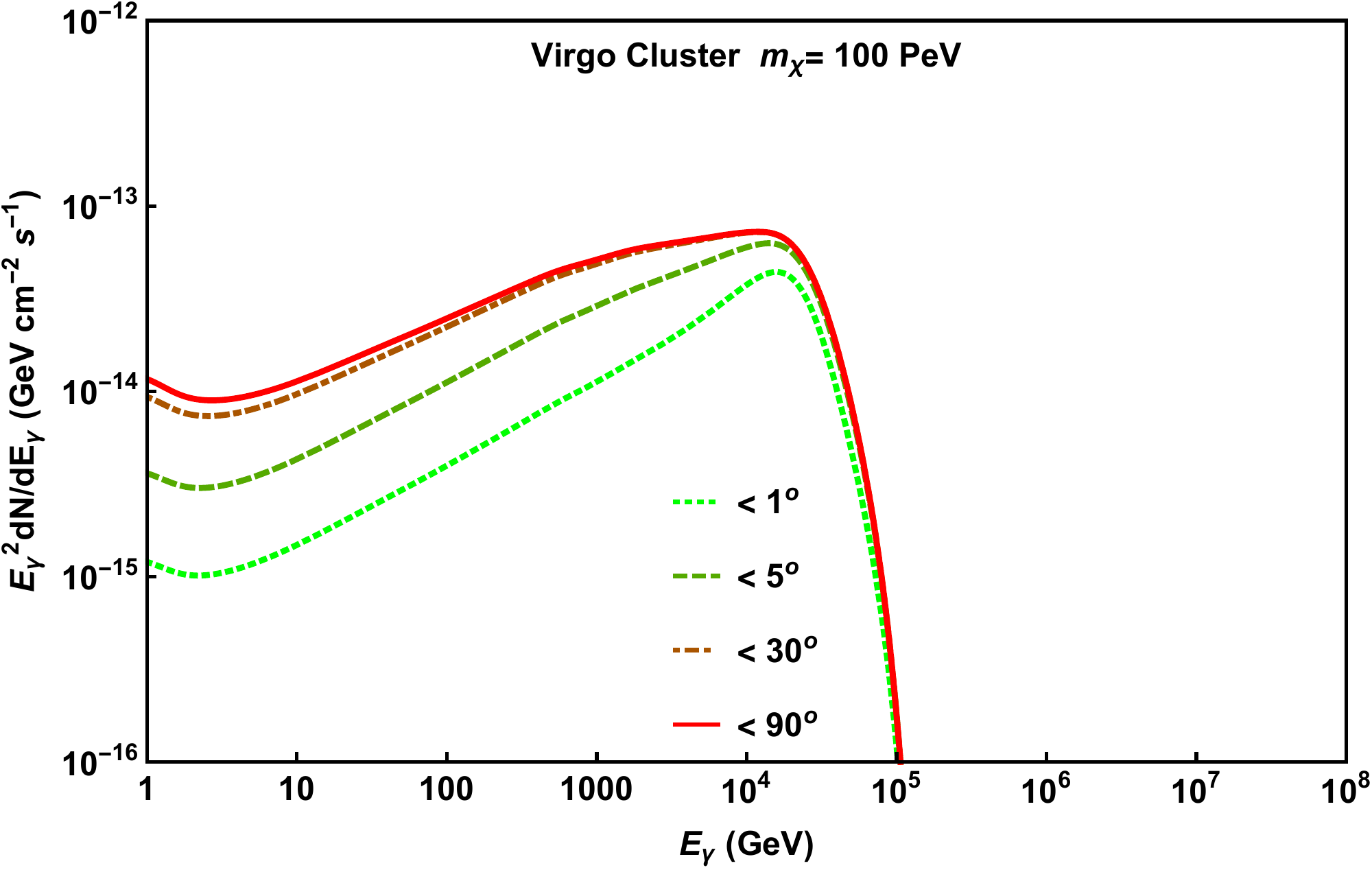}
\caption{The gamma-ray spectrum from the Virgo Cluster for the case of a dark matter particle with a mass of 0.1, 1, 10 or 100 PeV, annihilating to $b\bar{b}$ with a cross section of $\left\langle \sigma v\right\rangle =10^{-23} \, \rm{cm}^{3}/\rm{s}$. In each frame, we plot the spectrum integrated within a cone of radius $1^{\circ}$, $5^{\circ}$, $30^{\circ}$, or $90^{\circ}$ around Virgo. Comparing this to the results shown in Fig.~\ref{virgospec}, we see that much of the cascade emission from this source is distributed in a region of the sky several degrees in radius.}
\label{virgoangle}
\end{figure}

Unlike in the case of Draco, we need to consider gamma-ray attenuation that takes place within the volume of the Virgo Cluster itself. We do this using the equations given in Sec.~\ref{cascadespec} and following the procedure described in Ref.~\cite{Blanco:2017bgl}. Pair production within Virgo is again dominated by scattering with the CMB (the energy densities in the infrared, optical, and ultraviolet radiation fields within Virgo are approximately two orders of magnitude lower than that of the CMB~\cite{kaastra2008clusters}). Adopting an average magnetic field strength of 1 $\mu$G within the innermost 500 kpc of Virgo, we find that 92\% of the total energy in $E_e \ll 1$ PeV electrons goes into the production of lower energy gamma-rays via ICS. At higher energies, ICS is Klein-Nishina suppressed, and energy losses are instead dominated by synchrotron emission. 

After escaping from the Virgo cluster, gamma rays are further attenuated by the CMB, as well as by the cosmic infrared background (for which we adopt the model of Ref.~\cite{Dominguez:2010bv}). Synchrotron losses are negligible in this regime. In Fig.~\ref{virgospec}, we plot the gamma-ray spectrum from Virgo for a dark matter particle with mass of 0.1, 1, 10 or 100 PeV, annihilating to $b\bar{b}$ with a cross section of $\left\langle \sigma v\right\rangle =10^{-23} \, \rm{cm}^{3}/\rm{s}$. In each frame, the black dot-dashed curve denotes the injected spectrum, without taking into account any attenuation or cascades, while the dashed orange line represents the spectrum which escapes from the Virgo Cluster itself, including both the unattenuated gamma-rays and those generated in the subsequent cascade. The blue dotted (red solid) line is the spectrum that reaches the Solar System without (with) the contribution from cascades. Note that in the $m_{\chi}=100$ TeV case, the dashed orange curve (leaving Virgo) falls directly underneath the black dot-dashed curve (injected).

Within Virgo itself, the CMB strongly attenuates the gamma-ray spectrum above a few hundred TeV. Much of this energy is lost to synchrotron emission. After escaping Virgo, the CMB and cosmic infrared background further attenuate the gamma-ray spectrum over the course of its considerable 16.8 Mpc baseline.

In Fig.~\ref{virgoangle}, we plot the gamma-ray spectrum from Virgo, as integrated within a cone of radius $1^{\circ}$, $5^{\circ}$, $30^{\circ}$, or $90^{\circ}$. Here, we have adopted an extragalactic magnetic field of $B \sim 10^{-10}$ G, which is sufficient to isotropize the ICS emission from electrons $E_e \lsim 200$ TeV. Again, we find that much of the cascade emission from this source is expected to be distributed across a region of the sky several degrees in radius. Note that if we had assumed a lower value for the average extragalactic magnetic field, the cascade would be less extended on the sky, especially at high energies.

\section{Summary and Conclusions}

The universe is approximately transparent to gamma rays with energies up to $\sim$100 GeV over cosmological distances, and up to $\sim$100 TeV over Galactic distance scales. For this reason, most gamma-ray searches for dark matter in dwarf galaxies, the Galactic Center, and nearby galaxy clusters can safely neglect the impact of any interactions experienced during propagation. If the dark matter is approximately $\sim$100 TeV or heavier, however, the very high-energy annihilation or decay products can be significantly attenuated through the process of pair production with the cosmic microwave and infrared backgrounds. These electron-positron pairs will then be deflected by magnetic fields as they  transfer their energy through inverse Compton scattering, resulting in the formation and evolution of electromagnetic cascades. This alters not only the spectral shape of the gamma-ray signal, but also the angular profile of this emission. More specifically, we calculate that the annihilations or decays of a typical PeV-scale dark matter candidate will produce extended halos of gamma ray emission from dwarf galaxies and nearby galaxy clusters, several degrees in extent. The new ground-based gamma-ray observatory HAWC is well suited to search for and potentially detect these novel signatures of very massive dark matter particles. We note that the effects of attenuation and cascades described in this study are not limited to the case of very heavy dark matter particles, but instead represent universal signatures of distant PeV-scale gamma-ray sources.

\bigskip

\textbf{Acknowledgments.} CB is supported by the US National Science Foundation Graduate Research Fellowship under grants number DGE-1144082 and DGE-1746045. JPH is supported by the US Department of Energy Office of High-Energy Physics and the Laboratory Directed Research and Development (LDRD) program of Los Alamos National Laboratory. This manuscript has been authored by Fermi Research Alliance, LLC under Contract No. DE-AC02-07CH11359 with the U.S. Department of Energy, Office of Science, Office of High Energy Physics. The United States Government retains and the publisher, by accepting the article for publication, acknowledges that the United States Government retains a non-exclusive, paid-up, irrevocable, world-wide license to publish or reproduce the published form of this manuscript, or allow others to do so, for United States Government purposes.

\bibliography{HAWCbib}

\providecommand{\href}[2]{#2}\begingroup\raggedright\begin{thebibliography}{10}

\bibitem{Fermi-LAT:2016uux}
{\scshape DES, Fermi-LAT} collaboration, A.~Albert et~al., \emph{{Searching for
  Dark Matter Annihilation in Recently Discovered Milky Way Satellites with
  Fermi-LAT}},  \href{http://arxiv.org/abs/1611.03184}{{\tt 1611.03184}}.

\bibitem{Drlica-Wagner:2015xua}
{\scshape DES, Fermi-LAT} collaboration, A.~Drlica-Wagner et~al., \emph{{Search
  for Gamma-Ray Emission from DES Dwarf Spheroidal Galaxy Candidates with
  Fermi-LAT Data}},
  \href{http://dx.doi.org/10.1088/2041-8205/809/1/L4}{\emph{Astrophys. J.} {\bf
  809} (2015) L4}, [\href{http://arxiv.org/abs/1503.02632}{{\tt 1503.02632}}].

\bibitem{Geringer-Sameth:2014qqa}
A.~Geringer-Sameth, S.~M. Koushiappas and M.~G. Walker, \emph{{A Comprehensive
  Search for Dark Matter Annihilation in Dwarf Galaxies}},
  \href{http://arxiv.org/abs/1410.2242}{{\tt 1410.2242}}.

\bibitem{Albert:2017vtb}
{\scshape HAWC} collaboration, A.~Albert et~al., \emph{{Dark Matter Limits From
  Dwarf Spheroidal Galaxies with The HAWC Gamma-Ray Observatory}},
  \href{http://arxiv.org/abs/1706.01277}{{\tt 1706.01277}}.

\bibitem{TheFermi-LAT:2017vmf}
{\scshape Fermi-LAT} collaboration, M.~Ackermann et~al., \emph{{The Fermi
  Galactic Center GeV Excess and Implications for Dark Matter}},
  \href{http://dx.doi.org/10.3847/1538-4357/aa6cab}{\emph{Astrophys. J.} {\bf
  840} (2017) 43}, [\href{http://arxiv.org/abs/1704.03910}{{\tt 1704.03910}}].

\bibitem{Hooper:2012sr}
D.~Hooper, C.~Kelso and F.~S. Queiroz, \emph{{Stringent and Robust Constraints
  on the Dark Matter Annihilation Cross Section From the Region of the Galactic
  Center}},
  \href{http://dx.doi.org/10.1016/j.astropartphys.2013.04.007}{\emph{Astropart.
  Phys.} {\bf 46} (2013) 55--70}, [\href{http://arxiv.org/abs/1209.3015}{{\tt
  1209.3015}}].

\bibitem{Abeysekara:2014ffg}
{\scshape HAWC} collaboration, A.~U. Abeysekara et~al., \emph{{Sensitivity of
  HAWC to high-mass dark matter annihilations}},
  \href{http://dx.doi.org/10.1103/PhysRevD.90.122002}{\emph{Phys. Rev.} {\bf
  D90} (2014) 122002}, [\href{http://arxiv.org/abs/1405.1730}{{\tt
  1405.1730}}].

\bibitem{Lisanti:2017qlb}
M.~Lisanti, S.~Mishra-Sharma, N.~L. Rodd and B.~R. Safdi, \emph{{A Search for
  Dark Matter Annihilation in Galaxy Groups}},
  \href{http://arxiv.org/abs/1708.09385}{{\tt 1708.09385}}.

\bibitem{Ackermann:2015fdi}
{\scshape Fermi-LAT} collaboration, M.~Ackermann et~al., \emph{{Search for
  extended gamma-ray emission from the Virgo galaxy cluster with Fermi-LAT}},
  \href{http://dx.doi.org/10.1088/0004-637X/812/2/159}{\emph{Astrophys. J.}
  {\bf 812} (2015) 159}, [\href{http://arxiv.org/abs/1510.00004}{{\tt
  1510.00004}}].

\bibitem{Ackermann:2015tah}
{\scshape Fermi-LAT} collaboration, M.~Ackermann et~al., \emph{{Limits on Dark
  Matter Annihilation Signals from the Fermi LAT 4-year Measurement of the
  Isotropic Gamma-Ray Background}},
  \href{http://dx.doi.org/10.1088/1475-7516/2015/09/008}{\emph{JCAP} {\bf 1509}
  (2015) 008}, [\href{http://arxiv.org/abs/1501.05464}{{\tt 1501.05464}}].

\bibitem{DiMauro:2015tfa}
M.~Di~Mauro and F.~Donato, \emph{{Composition of the Fermi-LAT isotropic
  gamma-ray background intensity: Emission from extragalactic point sources and
  dark matter annihilations}},
  \href{http://dx.doi.org/10.1103/PhysRevD.91.123001}{\emph{Phys. Rev.} {\bf
  D91} (2015) 123001}, [\href{http://arxiv.org/abs/1501.05316}{{\tt
  1501.05316}}].

\bibitem{Griest:1989wd}
K.~Griest and M.~Kamionkowski, \emph{{Unitarity Limits on the Mass and Radius
  of Dark Matter Particles}},
  \href{http://dx.doi.org/10.1103/PhysRevLett.64.615}{\emph{Phys. Rev. Lett.}
  {\bf 64} (1990) 615}.

\bibitem{Berlin:2016vnh}
A.~Berlin, D.~Hooper and G.~Krnjaic, \emph{{PeV-Scale Dark Matter as a Thermal
  Relic of a Decoupled Sector}},
  \href{http://dx.doi.org/10.1016/j.physletb.2016.06.037}{\emph{Phys. Lett.}
  {\bf B760} (2016) 106--111}, [\href{http://arxiv.org/abs/1602.08490}{{\tt
  1602.08490}}].

\bibitem{Berlin:2016gtr}
A.~Berlin, D.~Hooper and G.~Krnjaic, \emph{{Thermal Dark Matter From A Highly
  Decoupled Sector}},
  \href{http://dx.doi.org/10.1103/PhysRevD.94.095019}{\emph{Phys. Rev.} {\bf
  D94} (2016) 095019}, [\href{http://arxiv.org/abs/1609.02555}{{\tt
  1609.02555}}].

\bibitem{Fornengo:2002db}
N.~Fornengo, A.~Riotto and S.~Scopel, \emph{{Supersymmetric dark matter and the
  reheating temperature of the universe}},
  \href{http://dx.doi.org/10.1103/PhysRevD.67.023514}{\emph{Phys. Rev.} {\bf
  D67} (2003) 023514}, [\href{http://arxiv.org/abs/hep-ph/0208072}{{\tt
  hep-ph/0208072}}].

\bibitem{Gelmini:2006pq}
G.~Gelmini, P.~Gondolo, A.~Soldatenko and C.~E. Yaguna, \emph{{The Effect of a
  late decaying scalar on the neutralino relic density}},
  \href{http://dx.doi.org/10.1103/PhysRevD.74.083514}{\emph{Phys. Rev.} {\bf
  D74} (2006) 083514}, [\href{http://arxiv.org/abs/hep-ph/0605016}{{\tt
  hep-ph/0605016}}].

\bibitem{Hooper:2013nia}
D.~Hooper, \emph{{Is the CMB Telling Us that Dark Matter is Weaker than Weakly
  Interacting?}},
  \href{http://dx.doi.org/10.1103/PhysRevD.88.083519}{\emph{Phys. Rev.} {\bf
  D88} (2013) 083519}, [\href{http://arxiv.org/abs/1307.0826}{{\tt
  1307.0826}}].

\bibitem{Kane:2015jia}
G.~Kane, K.~Sinha and S.~Watson, \emph{{Cosmological Moduli and the
  Post-Inflationary Universe: A Critical Review}},
  \href{http://dx.doi.org/10.1142/S0218271815300220}{\emph{Int. J. Mod. Phys.}
  {\bf D24} (2015) 1530022}, [\href{http://arxiv.org/abs/1502.07746}{{\tt
  1502.07746}}].

\bibitem{Patwardhan:2015kga}
A.~V. Patwardhan, G.~M. Fuller, C.~T. Kishimoto and A.~Kusenko, \emph{{Diluted
  equilibrium sterile neutrino dark matter}},
  \href{http://dx.doi.org/10.1103/PhysRevD.92.103509}{\emph{Phys. Rev.} {\bf
  D92} (2015) 103509}, [\href{http://arxiv.org/abs/1507.01977}{{\tt
  1507.01977}}].

\bibitem{Hoof:2017ibo}
S.~Hoof and J.~Jaeckel, \emph{{QCD axions and axion-like particles in a
  2-inflation scenario}},  \href{http://arxiv.org/abs/1709.01090}{{\tt
  1709.01090}}.

\bibitem{Bramante:2017obj}
J.~Bramante and J.~Unwin, \emph{{Superheavy Thermal Dark Matter and Primordial
  Asymmetries}}, \href{http://dx.doi.org/10.1007/JHEP02(2017)119}{\emph{JHEP}
  {\bf 02} (2017) 119}, [\href{http://arxiv.org/abs/1701.05859}{{\tt
  1701.05859}}].

\bibitem{Giblin:2017wlo}
J.~T. Giblin, G.~Kane, E.~Nesbit, S.~Watson and Y.~Zhao, \emph{{Was the
  Universe Actually Radiation Dominated Prior to Nucleosynthesis?}},
  \href{http://dx.doi.org/10.1103/PhysRevD.96.043525}{\emph{Phys. Rev.} {\bf
  D96} (2017) 043525}, [\href{http://arxiv.org/abs/1706.08536}{{\tt
  1706.08536}}].

\bibitem{Davoudiasl:2015vba}
H.~Davoudiasl, D.~Hooper and S.~D. McDermott, \emph{{Inflatable Dark Matter}},
  \href{http://dx.doi.org/10.1103/PhysRevLett.116.031303}{\emph{Phys. Rev.
  Lett.} {\bf 116} (2016) 031303}, [\href{http://arxiv.org/abs/1507.08660}{{\tt
  1507.08660}}].

\bibitem{Lyth:1995ka}
D.~H. Lyth and E.~D. Stewart, \emph{{Thermal inflation and the moduli
  problem}}, \href{http://dx.doi.org/10.1103/PhysRevD.53.1784}{\emph{Phys.
  Rev.} {\bf D53} (1996) 1784--1798},
  [\href{http://arxiv.org/abs/hep-ph/9510204}{{\tt hep-ph/9510204}}].

\bibitem{Boeckel:2011yj}
T.~Boeckel and J.~Schaffner-Bielich, \emph{{A little inflation at the
  cosmological QCD phase transition}},
  \href{http://dx.doi.org/10.1103/PhysRevD.85.103506}{\emph{Phys. Rev.} {\bf
  D85} (2012) 103506}, [\href{http://arxiv.org/abs/1105.0832}{{\tt
  1105.0832}}].

\bibitem{Boeckel:2009ej}
T.~Boeckel and J.~Schaffner-Bielich, \emph{{A little inflation in the early
  universe at the QCD phase transition}},
  \href{http://dx.doi.org/10.1103/PhysRevLett.105.041301,
  10.1103/PhysRevLett.106.069901}{\emph{Phys. Rev. Lett.} {\bf 105} (2010)
  041301}, [\href{http://arxiv.org/abs/0906.4520}{{\tt 0906.4520}}].

\bibitem{Chung:1998zb}
D.~J.~H. Chung, E.~W. Kolb and A.~Riotto, \emph{{Superheavy dark matter}},
  \href{http://dx.doi.org/10.1103/PhysRevD.59.023501}{\emph{Phys. Rev.} {\bf
  D59} (1999) 023501}, [\href{http://arxiv.org/abs/hep-ph/9802238}{{\tt
  hep-ph/9802238}}].

\bibitem{Chung:1998rq}
D.~J.~H. Chung, E.~W. Kolb and A.~Riotto, \emph{{Production of massive
  particles during reheating}},
  \href{http://dx.doi.org/10.1103/PhysRevD.60.063504}{\emph{Phys. Rev.} {\bf
  D60} (1999) 063504}, [\href{http://arxiv.org/abs/hep-ph/9809453}{{\tt
  hep-ph/9809453}}].

\bibitem{Allahverdi:2010xz}
R.~Allahverdi, R.~Brandenberger, F.-Y. Cyr-Racine and A.~Mazumdar,
  \emph{{Reheating in Inflationary Cosmology: Theory and Applications}},
  \href{http://dx.doi.org/10.1146/annurev.nucl.012809.104511}{\emph{Ann. Rev.
  Nucl. Part. Sci.} {\bf 60} (2010) 27--51},
  [\href{http://arxiv.org/abs/1001.2600}{{\tt 1001.2600}}].

\bibitem{Giudice:1999yt}
G.~F. Giudice, I.~Tkachev and A.~Riotto, \emph{{Nonthermal production of
  dangerous relics in the early universe}},
  \href{http://dx.doi.org/10.1088/1126-6708/1999/08/009}{\emph{JHEP} {\bf 08}
  (1999) 009}, [\href{http://arxiv.org/abs/hep-ph/9907510}{{\tt
  hep-ph/9907510}}].

\bibitem{Chung:1998ua}
D.~J.~H. Chung, E.~W. Kolb and A.~Riotto, \emph{{Nonthermal supermassive dark
  matter}}, \href{http://dx.doi.org/10.1103/PhysRevLett.81.4048}{\emph{Phys.
  Rev. Lett.} {\bf 81} (1998) 4048--4051},
  [\href{http://arxiv.org/abs/hep-ph/9805473}{{\tt hep-ph/9805473}}].

\bibitem{Chianese:2017nwe}
M.~Chianese, G.~Miele and S.~Morisi, \emph{{Interpreting IceCube 6-year HESE
  data as an evidence for hundred TeV decaying Dark Matter}},
  \href{http://dx.doi.org/10.1016/j.physletb.2017.09.016}{\emph{Phys. Lett.}
  {\bf B773} (2017) 591--595}, [\href{http://arxiv.org/abs/1707.05241}{{\tt
  1707.05241}}].

\bibitem{Sahoo:2017cqg}
B.~Sahoo, M.~K. Parida and M.~Chakraborty, \emph{{A Benchmark Model from
  SO(10): Dark Matter Decay for IceCube Neutrinos and Verifiable Proton
  Decay}},  \href{http://arxiv.org/abs/1707.01286}{{\tt 1707.01286}}.

\bibitem{Bhattacharya:2017jaw}
A.~Bhattacharya, A.~Esmaili, S.~Palomares-Ruiz and I.~Sarcevic, \emph{{Probing
  decaying heavy dark matter with the 4-year IceCube HESE data}},
  \href{http://dx.doi.org/10.1088/1475-7516/2017/07/027}{\emph{JCAP} {\bf 1707}
  (2017) 027}, [\href{http://arxiv.org/abs/1706.05746}{{\tt 1706.05746}}].

\bibitem{Borah:2017xgm}
D.~Borah, A.~Dasgupta, U.~K. Dey, S.~Patra and G.~Tomar, \emph{{Multi-component
  Fermionic Dark Matter and IceCube PeV scale Neutrinos in Left-Right Model
  with Gauge Unification}},
  \href{http://dx.doi.org/10.1007/JHEP09(2017)005}{\emph{JHEP} {\bf 09} (2017)
  005}, [\href{http://arxiv.org/abs/1704.04138}{{\tt 1704.04138}}].

\bibitem{Cohen:2016uyg}
T.~Cohen, K.~Murase, N.~L. Rodd, B.~R. Safdi and Y.~Soreq, \emph{{γ -ray
  Constraints on Decaying Dark Matter and Implications for IceCube}},
  \href{http://dx.doi.org/10.1103/PhysRevLett.119.021102}{\emph{Phys. Rev.
  Lett.} {\bf 119} (2017) 021102}, [\href{http://arxiv.org/abs/1612.05638}{{\tt
  1612.05638}}].

\bibitem{Chianese:2016kpu}
M.~Chianese, G.~Miele and S.~Morisi, \emph{{Dark Matter interpretation of low
  energy IceCube MESE excess}},
  \href{http://dx.doi.org/10.1088/1475-7516/2017/01/007}{\emph{JCAP} {\bf 1701}
  (2017) 007}, [\href{http://arxiv.org/abs/1610.04612}{{\tt 1610.04612}}].

\bibitem{Chianese:2016smc}
M.~Chianese and A.~Merle, \emph{{A Consistent Theory of Decaying Dark Matter
  Connecting IceCube to the Sesame Street}},
  \href{http://dx.doi.org/10.1088/1475-7516/2017/04/017}{\emph{JCAP} {\bf 1704}
  (2017) 017}, [\href{http://arxiv.org/abs/1607.05283}{{\tt 1607.05283}}].

\bibitem{Dev:2016qbd}
P.~S.~B. Dev, D.~Kazanas, R.~N. Mohapatra, V.~L. Teplitz and Y.~Zhang,
  \emph{{Heavy right-handed neutrino dark matter and PeV neutrinos at
  IceCube}}, \href{http://dx.doi.org/10.1088/1475-7516/2016/08/034}{\emph{JCAP}
  {\bf 1608} (2016) 034}, [\href{http://arxiv.org/abs/1606.04517}{{\tt
  1606.04517}}].

\bibitem{Fiorentin:2016avj}
M.~Re~Fiorentin, V.~Niro and N.~Fornengo, \emph{{A consistent model for
  leptogenesis, dark matter and the IceCube signal}},
  \href{http://dx.doi.org/10.1007/JHEP11(2016)022}{\emph{JHEP} {\bf 11} (2016)
  022}, [\href{http://arxiv.org/abs/1606.04445}{{\tt 1606.04445}}].

\bibitem{Chianese:2016tmd}
M.~Chianese, \emph{{IceCube PeV Neutrinos and Leptophilic Dark Matter}},
  \href{http://dx.doi.org/10.1088/1742-6596/718/4/042014}{\emph{J. Phys. Conf.
  Ser.} {\bf 718} (2016) 042014}, [\href{http://arxiv.org/abs/1605.05749}{{\tt
  1605.05749}}].

\bibitem{Chianese:2016opp}
M.~Chianese, G.~Miele, S.~Morisi and E.~Vitagliano, \emph{{Low energy IceCube
  data and a possible Dark Matter related excess}},
  \href{http://dx.doi.org/10.1016/j.physletb.2016.03.084}{\emph{Phys. Lett.}
  {\bf B757} (2016) 251--256}, [\href{http://arxiv.org/abs/1601.02934}{{\tt
  1601.02934}}].

\bibitem{Biondi:2015yia}
R.~Biondi, \emph{{Dark Matter and IceCube Neutrinos}},
  \href{http://dx.doi.org/10.1393/ncc/i2015-15031-4}{\emph{Nuovo Cim.} {\bf
  C38} (2015) 31}, [\href{http://arxiv.org/abs/1508.06089}{{\tt 1508.06089}}].

\bibitem{Boucenna:2015tra}
S.~M. Boucenna, M.~Chianese, G.~Mangano, G.~Miele, S.~Morisi, O.~Pisanti
  et~al., \emph{{Decaying Leptophilic Dark Matter at IceCube}},
  \href{http://dx.doi.org/10.1088/1475-7516/2015/12/055}{\emph{JCAP} {\bf 1512}
  (2015) 055}, [\href{http://arxiv.org/abs/1507.01000}{{\tt 1507.01000}}].

\bibitem{Berezhiani:2015fba}
Z.~Berezhiani, \emph{{Shadow dark matter, sterile neutrinos and neutrino events
  at IceCube}},
  \href{http://dx.doi.org/10.1016/j.nuclphysbps.2015.06.076}{\emph{Nucl. Part.
  Phys. Proc.} {\bf 265-266} (2015) 303--306},
  [\href{http://arxiv.org/abs/1506.09040}{{\tt 1506.09040}}].

\bibitem{Anchordoqui:2015lqa}
L.~A. Anchordoqui, V.~Barger, H.~Goldberg, X.~Huang, D.~Marfatia, L.~H.~M.
  da~Silva et~al., \emph{{IceCube neutrinos, decaying dark matter, and the
  Hubble constant}}, \href{http://dx.doi.org/10.1103/PhysRevD.92.061301,
  10.1103/PhysRevD.94.069901}{\emph{Phys. Rev.} {\bf D92} (2015) 061301},
  [\href{http://arxiv.org/abs/1506.08788}{{\tt 1506.08788}}].

\bibitem{Murase:2015gea}
K.~Murase, R.~Laha, S.~Ando and M.~Ahlers, \emph{{Testing the Dark Matter
  Scenario for PeV Neutrinos Observed in IceCube}},
  \href{http://dx.doi.org/10.1103/PhysRevLett.115.071301}{\emph{Phys. Rev.
  Lett.} {\bf 115} (2015) 071301}, [\href{http://arxiv.org/abs/1503.04663}{{\tt
  1503.04663}}].

\bibitem{Esmaili:2014rma}
A.~Esmaili, S.~K. Kang and P.~D. Serpico, \emph{{IceCube events and decaying
  dark matter: hints and constraints}},
  \href{http://dx.doi.org/10.1088/1475-7516/2014/12/054}{\emph{JCAP} {\bf 1412}
  (2014) 054}, [\href{http://arxiv.org/abs/1410.5979}{{\tt 1410.5979}}].

\bibitem{Rott:2014kfa}
C.~Rott, K.~Kohri and S.~C. Park, \emph{{Superheavy dark matter and IceCube
  neutrino signals: Bounds on decaying dark matter}},
  \href{http://dx.doi.org/10.1103/PhysRevD.92.023529}{\emph{Phys. Rev.} {\bf
  D92} (2015) 023529}, [\href{http://arxiv.org/abs/1408.4575}{{\tt
  1408.4575}}].

\bibitem{Zavala:2014dla}
J.~Zavala, \emph{{Galactic PeV neutrinos from dark matter annihilation}},
  \href{http://dx.doi.org/10.1103/PhysRevD.89.123516}{\emph{Phys. Rev.} {\bf
  D89} (2014) 123516}, [\href{http://arxiv.org/abs/1404.2932}{{\tt
  1404.2932}}].

\bibitem{Bhattacharya:2014vwa}
A.~Bhattacharya, M.~H. Reno and I.~Sarcevic, \emph{{Reconciling neutrino flux
  from heavy dark matter decay and recent events at IceCube}},
  \href{http://dx.doi.org/10.1007/JHEP06(2014)110}{\emph{JHEP} {\bf 06} (2014)
  110}, [\href{http://arxiv.org/abs/1403.1862}{{\tt 1403.1862}}].

\bibitem{Bai:2013nga}
Y.~Bai, R.~Lu and J.~Salvado, \emph{{Geometric Compatibility of IceCube TeV-PeV
  Neutrino Excess and its Galactic Dark Matter Origin}},
  \href{http://dx.doi.org/10.1007/JHEP01(2016)161}{\emph{JHEP} {\bf 01} (2016)
  161}, [\href{http://arxiv.org/abs/1311.5864}{{\tt 1311.5864}}].

\bibitem{Esmaili:2013gha}
A.~Esmaili and P.~D. Serpico, \emph{{Are IceCube neutrinos unveiling PeV-scale
  decaying dark matter?}},
  \href{http://dx.doi.org/10.1088/1475-7516/2013/11/054}{\emph{JCAP} {\bf 1311}
  (2013) 054}, [\href{http://arxiv.org/abs/1308.1105}{{\tt 1308.1105}}].

\bibitem{Harding:2015bua}
{\scshape HAWC} collaboration, J.~P. Harding and B.~Dingus, \emph{{Dark Matter
  Annihilation and Decay Searches with the High Altitude Water Cherenkov (HAWC)
  Observatory}}, {\emph{PoS} {\bf ICRC2015} (2016) 1227},
  [\href{http://arxiv.org/abs/1508.04352}{{\tt 1508.04352}}].

\bibitem{Proper:2015xya}
{\scshape HAWC} collaboration, M.~L. Proper, J.~P. Harding and B.~Dingus,
  \emph{{First Limits on the Dark Matter Cross Section with the HAWC
  Observatory}}, {\emph{PoS} {\bf ICRC2015} (2016) 1213},
  [\href{http://arxiv.org/abs/1508.04470}{{\tt 1508.04470}}].

\bibitem{Yapici:2017bxr}
{\scshape HAWC} collaboration, T.~Yapici and A.~J. Smith, \emph{{Dark Matter
  Searches with HAWC}},  in \emph{{Proceedings, 35th International Cosmic Ray
  Conference (ICRC 2017): Bexco, Busan, Korea, July 12-20, 2017}}, 2017.
\newblock \href{http://arxiv.org/abs/1708.07461}{{\tt 1708.07461}}.

\bibitem{Cadena:2017jmw}
{\scshape HAWC} collaboration, S.~H. Cadena, \emph{{Searching dark matter
  signatures from the Virgo cluster with HAWC}},
  \href{http://dx.doi.org/10.1063/1.4968993}{\emph{AIP Conf. Proc.} {\bf 1792}
  (2017) 060010}.

\bibitem{Abeysekara:2017jxs}
A.~U. Abeysekara et~al., \emph{{A Search for Dark Matter in the Galactic Halo
  with HAWC}},  \href{http://arxiv.org/abs/1710.10288}{{\tt 1710.10288}}.

\bibitem{Abeysekara:2017mjj}
A.~U. Abeysekara et~al., \emph{{Observation of the Crab Nebula with the HAWC
  Gamma-Ray Observatory}},
  \href{http://dx.doi.org/10.3847/1538-4357/aa7555}{\emph{Astrophys. J.} {\bf
  843} (2017) 39}, [\href{http://arxiv.org/abs/1701.01778}{{\tt 1701.01778}}].

\bibitem{Abeysekara:2017hyn}
A.~U. Abeysekara et~al., \emph{{The 2HWC HAWC Observatory Gamma Ray Catalog}},
  \href{http://dx.doi.org/10.3847/1538-4357/aa7556}{\emph{Astrophys. J.} {\bf
  843} (2017) 40}, [\href{http://arxiv.org/abs/1702.02992}{{\tt 1702.02992}}].

\bibitem{2012ApJ...750...63A}
A.~A. {Abdo} et~al., \emph{{Observation and Spectral Measurements of the Crab
  Nebula with Milagro}},
  \href{http://dx.doi.org/10.1088/0004-637X/750/1/63}{\emph{Astrophys. J} {\bf
  750} (May, 2012) 63}, [\href{http://arxiv.org/abs/1110.0409}{{\tt
  1110.0409}}].

\bibitem{Holler:2015uca}
{\scshape H.E.S.S.} collaboration, M.~Holler et~al., \emph{{Observations of the
  Crab Nebula with H.E.S.S. Phase II}}, {\emph{PoS} {\bf ICRC2015} (2016) 847},
  [\href{http://arxiv.org/abs/1509.02902}{{\tt 1509.02902}}].

\bibitem{2016APh....72...76A}
{Aleksi{\'c}} et~al., \emph{{The major upgrade of the MAGIC telescopes, Part
  II: A performance study using observations of the Crab Nebula}},
  \href{http://dx.doi.org/10.1016/j.astropartphys.2015.02.005}{\emph{Astroparticle
  Physics} {\bf 72} (Jan., 2016) 76--94},
  [\href{http://arxiv.org/abs/1409.5594}{{\tt 1409.5594}}].

\bibitem{2015ICRC...34..771P}
N.~{Park} and {VERITAS Collaboration}, \emph{{Performance of the VERITAS
  experiment}},  in \emph{34th International Cosmic Ray Conference (ICRC2015)},
  vol.~34 of \emph{International Cosmic Ray Conference}, p.~771, July, 2015.
\newblock \href{http://arxiv.org/abs/1508.07070}{{\tt 1508.07070}}.

\bibitem{Pass8}
https://www.slac.stanford.edu/exp/glast/groups/canda/lat Performance.htm.

\bibitem{Capistran:2017fxl}
{\scshape HAWC} collaboration, T.~Capistrán, I.~Torres and E.~Moreno,
  \emph{{Stability and behavior of the outer array of small water Cherenkov
  detectors, outriggers, in the HAWC observatory}},  in \emph{{Proceedings,
  35th International Cosmic Ray Conference (ICRC 2017): Bexco, Busan, Korea,
  July 12-20, 2017}}, 2017.
\newblock \href{http://arxiv.org/abs/1708.03598}{{\tt 1708.03598}}.

\bibitem{Joshi:2017eou}
{\scshape HAWC} collaboration, V.~Joshi and A.~Jardin-Blicq, \emph{{HAWC High
  Energy Upgrade with a Sparse Outrigger Array}},  in \emph{{Proceedings, 35th
  International Cosmic Ray Conference (ICRC 2017): Bexco, Busan, Korea, July
  12-20, 2017}}, 2017.
\newblock \href{http://arxiv.org/abs/1708.04032}{{\tt 1708.04032}}.

\bibitem{Marinelli:2017vzu}
{\scshape HAWC} collaboration, S.~S. Marinelli and J.~Goodman, \emph{{Measuring
  High-Energy Spectra with HAWC}},  in \emph{{Proceedings, 35th International
  Cosmic Ray Conference (ICRC 2017): Bexco, Busan, Korea, July 12-20, 2017}},
  2017.
\newblock \href{http://arxiv.org/abs/1708.03502}{{\tt 1708.03502}}.

\bibitem{Blanco:2017bgl}
C.~Blanco and D.~Hooper, \emph{{High-Energy Gamma Rays and Neutrinos from
  Nearby Radio Galaxies}},  \href{http://arxiv.org/abs/1706.07047}{{\tt
  1706.07047}}.

\bibitem{Murase:2011cy}
K.~Murase, C.~D. Dermer, H.~Takami and G.~Migliori, \emph{{Blazars as
  Ultra-High-Energy Cosmic-Ray Sources: Implications for TeV Gamma-Ray
  Observations}},
  \href{http://dx.doi.org/10.1088/0004-637X/749/1/63}{\emph{Astrophys. J.} {\bf
  749} (2012) 63}, [\href{http://arxiv.org/abs/1107.5576}{{\tt 1107.5576}}].

\bibitem{Murase:2011yw}
K.~Murase, \emph{{High-Energy Emission Induced by Ultra-high-Energy Photons as
  a Probe of Ultra-high-Energy Cosmic-Ray Accelerators Embedded in the Cosmic
  Web}}, \href{http://dx.doi.org/10.1088/2041-8205/745/2/L16}{\emph{Astrophys.
  J.} {\bf 745} (2012) L16}, [\href{http://arxiv.org/abs/1111.0936}{{\tt
  1111.0936}}].

\bibitem{Murase:2012df}
K.~Murase, J.~F. Beacom and H.~Takami, \emph{{Gamma-Ray and Neutrino
  Backgrounds as Probes of the High-Energy Universe: Hints of Cascades, General
  Constraints, and Implications for TeV Searches}},
  \href{http://dx.doi.org/10.1088/1475-7516/2012/08/030}{\emph{JCAP} {\bf 1208}
  (2012) 030}, [\href{http://arxiv.org/abs/1205.5755}{{\tt 1205.5755}}].

\bibitem{Murase:2012xs}
K.~Murase and J.~F. Beacom, \emph{{Constraining Very Heavy Dark Matter Using
  Diffuse Backgrounds of Neutrinos and Cascaded Gamma Rays}},
  \href{http://dx.doi.org/10.1088/1475-7516/2012/10/043}{\emph{JCAP} {\bf 1210}
  (2012) 043}, [\href{http://arxiv.org/abs/1206.2595}{{\tt 1206.2595}}].

\bibitem{Berezinsky:2016feh}
V.~Berezinsky and O.~Kalashev, \emph{{High energy electromagnetic cascades in
  extragalactic space: physics and features}},
  \href{http://dx.doi.org/10.1103/PhysRevD.94.023007}{\emph{Phys. Rev.} {\bf
  D94} (2016) 023007}, [\href{http://arxiv.org/abs/1603.03989}{{\tt
  1603.03989}}].

\bibitem{aharonian1983}
F.~Aharonian, A.~Atoian and A.~Nagapetian, \emph{Photoproduction of
  electron-positron pairs in compact x-ray sources}, {\emph{Astrofizika} {\bf
  19} (1983) 323--334}.

\bibitem{Dominguez:2010bv}
A.~Dominguez et~al., \emph{{Extragalactic Background Light Inferred from AEGIS
  Galaxy SED-type Fractions}},
  \href{http://dx.doi.org/10.1111/j.1365-2966.2010.17631.x}{\emph{Mon. Not.
  Roy. Astron. Soc.} {\bf 410} (2011) 2556},
  [\href{http://arxiv.org/abs/1007.1459}{{\tt 1007.1459}}].

\bibitem{aharonian1981}
F.~Aharonian and A.~Atoyan, \emph{Compton scattering of relativistic electrons
  in compact x-ray sources}, {\emph{Astrophysics and Space Science} {\bf 79}
  (1981) 321--336}.

\bibitem{Neronov:2009gh}
A.~Neronov and D.~V. Semikoz, \emph{{Sensitivity of gamma-ray telescopes for
  detection of magnetic fields in intergalactic medium}},
  \href{http://dx.doi.org/10.1103/PhysRevD.80.123012}{\emph{Phys. Rev.} {\bf
  D80} (2009) 123012}, [\href{http://arxiv.org/abs/0910.1920}{{\tt
  0910.1920}}].

\bibitem{Tashiro:2013bxa}
H.~Tashiro and T.~Vachaspati, \emph{{Cosmological magnetic field correlators
  from blazar induced cascade}},
  \href{http://dx.doi.org/10.1103/PhysRevD.87.123527}{\emph{Phys. Rev.} {\bf
  D87} (2013) 123527}, [\href{http://arxiv.org/abs/1305.0181}{{\tt
  1305.0181}}].

\bibitem{chen2015search}
W.~Chen, J.~H. Buckley and F.~Ferrer, \emph{Search for gev $\gamma$-ray pair
  halos around low redshift blazars}, {\emph{Physical review letters} {\bf 115}
  (2015) 211103}.

\bibitem{Abramowski:2014tra}
{\scshape H.E.S.S.} collaboration, A.~Abramowski et~al., \emph{{Search for dark
  matter annihilation signatures in H.E.S.S. observations of Dwarf Spheroidal
  Galaxies}}, \href{http://dx.doi.org/10.1103/PhysRevD.90.112012}{\emph{Phys.
  Rev.} {\bf D90} (2014) 112012}, [\href{http://arxiv.org/abs/1410.2589}{{\tt
  1410.2589}}].

\bibitem{Abramowski:2010aa}
{\scshape H.E.S.S.} collaboration, A.~Abramowski et~al., \emph{{H.E.S.S.
  constraints on Dark Matter annihilations towards the Sculptor and Carina
  Dwarf Galaxies}},
  \href{http://dx.doi.org/10.1016/j.astropartphys.2010.12.006}{\emph{Astropart.
  Phys.} {\bf 34} (2011) 608--616}, [\href{http://arxiv.org/abs/1012.5602}{{\tt
  1012.5602}}].

\bibitem{Archambault:2017wyh}
{\scshape VERITAS} collaboration, S.~Archambault et~al., \emph{{Dark Matter
  Constraints from a Joint Analysis of Dwarf Spheroidal Galaxy Observations
  with VERITAS}},
  \href{http://dx.doi.org/10.1103/PhysRevD.95.082001}{\emph{Phys. Rev.} {\bf
  D95} (2017) 082001}, [\href{http://arxiv.org/abs/1703.04937}{{\tt
  1703.04937}}].

\bibitem{Ahnen:2016qkx}
{\scshape Fermi-LAT, MAGIC} collaboration, M.~L. Ahnen et~al., \emph{{Limits to
  dark matter annihilation cross-section from a combined analysis of MAGIC and
  Fermi-LAT observations of dwarf satellite galaxies}},
  \href{http://dx.doi.org/10.1088/1475-7516/2016/02/039}{\emph{JCAP} {\bf 1602}
  (2016) 039}, [\href{http://arxiv.org/abs/1601.06590}{{\tt 1601.06590}}].

\bibitem{Aleksic:2011jx}
{\scshape MAGIC} collaboration, J.~Aleksic et~al., \emph{{Searches for Dark
  Matter annihilation signatures in the Segue 1 satellite galaxy with the
  MAGIC-I telescope}},
  \href{http://dx.doi.org/10.1088/1475-7516/2011/06/035}{\emph{JCAP} {\bf 1106}
  (2011) 035}, [\href{http://arxiv.org/abs/1103.0477}{{\tt 1103.0477}}].

\bibitem{pythia}
T.~Sjöstrand, S.~Ask, J.~R. Christiansen, R.~Corke, N.~Desai et~al., \emph{{An
  Introduction to PYTHIA 8.2}},  \href{http://arxiv.org/abs/1410.3012}{{\tt
  1410.3012}}.

\bibitem{Martinez:2013els}
G.~D. Martinez, \emph{{A Robust Determination of Milky Way Satellite Properties
  using Hierarchical Mass Modeling}},
  \href{http://arxiv.org/abs/1309.2641}{{\tt 1309.2641}}.

\bibitem{Geringer-Sameth:2014yza}
A.~Geringer-Sameth, S.~M. Koushiappas and M.~Walker, \emph{{Dwarf galaxy
  annihilation and decay emission profiles for dark matter experiments}},
  \href{http://dx.doi.org/10.1088/0004-637X/801/2/74}{\emph{Astrophys.J.} {\bf
  801} (2015) 74}, [\href{http://arxiv.org/abs/1408.0002}{{\tt 1408.0002}}].

\bibitem{Bonnivard:2015xpq}
V.~Bonnivard et~al., \emph{{Dark matter annihilation and decay in dwarf
  spheroidal galaxies: The classical and ultrafaint dSphs}},
  \href{http://dx.doi.org/10.1093/mnras/stv1601}{\emph{Mon. Not. Roy. Astron.
  Soc.} {\bf 453} (2015) 849--867},
  [\href{http://arxiv.org/abs/1504.02048}{{\tt 1504.02048}}].

\bibitem{Chiappo:2016xfs}
A.~Chiappo, J.~Cohen-Tanugi, J.~Conrad, L.~E. Strigari, B.~Anderson and M.~A.
  Sanchez-Conde, \emph{{Dwarf spheroidal J-factors without priors: A
  likelihood-based analysis for indirect dark matter searches}},
  \href{http://dx.doi.org/10.1093/mnras/stw3079}{\emph{Mon. Not. Roy. Astron.
  Soc.} {\bf 466} (2017) 669--676},
  [\href{http://arxiv.org/abs/1608.07111}{{\tt 1608.07111}}].

\bibitem{Ichikawa:2016nbi}
K.~Ichikawa, M.~N. Ishigaki, S.~Matsumoto, M.~Ibe, H.~Sugai, K.~Hayashi et~al.,
  \emph{{Foreground effect on the $J$-factor estimation of classical dwarf
  spheroidal galaxies}},
  \href{http://dx.doi.org/10.1093/mnras/stx682}{\emph{Mon. Not. Roy. Astron.
  Soc.} {\bf 468} (2017) 2884--2896},
  [\href{http://arxiv.org/abs/1608.01749}{{\tt 1608.01749}}].

\bibitem{Hayashi:2016kcy}
K.~Hayashi, K.~Ichikawa, S.~Matsumoto, M.~Ibe, M.~N. Ishigaki and H.~Sugai,
  \emph{{Dark matter annihilation and decay from non-spherical dark halos in
  galactic dwarf satellites}},
  \href{http://dx.doi.org/10.1093/mnras/stw1457}{\emph{Mon. Not. Roy. Astron.
  Soc.} {\bf 461} (2016) 2914--2928},
  [\href{http://arxiv.org/abs/1603.08046}{{\tt 1603.08046}}].

\bibitem{Ullio:2016kvy}
P.~Ullio and M.~Valli, \emph{{A critical reassessment of particle Dark Matter
  limits from dwarf satellites}},
  \href{http://dx.doi.org/10.1088/1475-7516/2016/07/025}{\emph{JCAP} {\bf 1607}
  (2016) 025}, [\href{http://arxiv.org/abs/1603.07721}{{\tt 1603.07721}}].

\bibitem{Abramowski:2012au}
{\scshape H.E.S.S.} collaboration, A.~Abramowski et~al., \emph{{Search for Dark
  Matter Annihilation Signals from the Fornax Galaxy Cluster with H.E.S.S}},
  \href{http://dx.doi.org/10.1088/0004-637X/750/2/123,
  10.1088/0004-637X/783/1/63}{\emph{Astrophys. J.} {\bf 750} (2012) 123},
  [\href{http://arxiv.org/abs/1202.5494}{{\tt 1202.5494}}].

\bibitem{Pfrommer:2012mm}
{\scshape VERITAS} collaboration, T.~Arlen et~al., \emph{{Constraints on Cosmic
  Rays, Magnetic Fields, and Dark Matter from Gamma-Ray Observations of the
  Coma Cluster of Galaxies with VERITAS and Fermi}},
  \href{http://dx.doi.org/10.1088/0004-637X/757/2/123}{\emph{Astrophys. J.}
  {\bf 757} (2012) 123}, [\href{http://arxiv.org/abs/1208.0676}{{\tt
  1208.0676}}].

\bibitem{Aleksic:2009ir}
{\scshape MAGIC} collaboration, J.~Aleksic et~al., \emph{{MAGIC Gamma-Ray
  Telescope Observation of the Perseus Cluster of Galaxies: Implications for
  Cosmic Rays, Dark Matter and NGC 1275}},
  \href{http://dx.doi.org/10.1088/0004-637X/710/1/634}{\emph{Astrophys. J.}
  {\bf 710} (2010) 634--647}, [\href{http://arxiv.org/abs/0909.3267}{{\tt
  0909.3267}}].

\bibitem{Navarro:1995iw}
J.~F. Navarro, C.~S. Frenk and S.~D. White, \emph{{The Structure of cold dark
  matter halos}}, \href{http://dx.doi.org/10.1086/177173}{\emph{Astrophys.J.}
  {\bf 462} (1996) 563--575},
  [\href{http://arxiv.org/abs/astro-ph/9508025}{{\tt astro-ph/9508025}}].

\bibitem{han2012constraining}
J.~Han, C.~S. Frenk, V.~R. Eke, L.~Gao, S.~D. White, A.~Boyarsky et~al.,
  \emph{Constraining extended gamma-ray emission from galaxy clusters},
  {\emph{Monthly Notices of the Royal Astronomical Society} {\bf 427} (2012)
  1651--1665}.

\bibitem{kaastra2008clusters}
J.~S. Kaastra, A.~Bykov, S.~Schindler, J.~Bleeker, S.~Borgani, A.~Diaferio
  et~al., \emph{Clusters of galaxies: beyond the thermal view}.
\newblock Springer, 2008.

\end{thebibliography}\endgroup
 \bibliographystyle{JHEP}

\end{document}